\documentclass[fleqn,usenatbib]{mnras}

\usepackage{newtxtext,newtxmath}

\usepackage[T1]{fontenc}
\usepackage{ae,aecompl}

\usepackage{etoolbox}

\usepackage{graphicx}	
\usepackage{amsmath}	
\usepackage{amssymb}	

\title[Common envelope ejection timescales]{Inferred timescales for common envelope ejection using wide astrometric companions}

\author[A.P. Igoshev, H.B. Perets \& E. Michaely]{
Andrei P. Igoshev $^{1,2}$\thanks{E-mail: ignotur@gmail.com}
Hagai B. Perets$^{1}$ \&
Erez Michaely$^{3}$
\\
$^{1}$ Physics Department, Technion - Israel Institute of Technology, Haifa 3200002, Israel \\
$^{2}$ Department of Applied Mathematics, University of Leeds, Leeds LS2 9JT, UK \\
$^{3}$ Astronomy Department, University of Maryland, College Park, MD 20742, USA
}

\date{Accepted XXX. Received YYY; in original form ZZZ}

\pubyear{2015}

\begin{document}
\label{firstpage}
\pagerange{\pageref{firstpage}--\pageref{lastpage}}
\maketitle

\begin{abstract}
Evolution of close binaries often proceeds through the common envelope stage. The physics of the envelope ejection (CEE) is not yet understood,  and several mechanisms were suggested to be involved. These could give rise to different timescales for the CEE mass-loss. 
In order to probe the CEE-timescales we study wide companions to post-CE binaries. Faster mass-loss timescales give rise to higher disruption rates of wide binaries and result in larger average separations. 
We make use of data from Gaia DR2 to search for ultra-wide companions (projected separations $10^3$~--~$2\times 10^5$~a.u. and $M_2 > 0.4$~M$_\odot$) to several types of post-CEE systems, including sdBs, white-dwarf post-common binaries, and cataclysmic variables. We find a (wide-orbit) multiplicity fraction of $1.4\pm 0.2$~per cent for sdBs to be compared with a multiplicity fraction of $5.0\pm 0.2$~per cent for late-B/A/F stars which are possible sdB progenitors. The distribution of projected separations of ultra-wide pairs to main sequence stars and sdBs differs significantly and is compatible with prompt mass loss (upper limit on common envelope ejection timescale of $10^2$ years). The smaller statistics of ultra-wide companions to cataclysmic variables and post-CEE binaries provide weaker constraints. Nevertheless, the survival rate of ultra-wide pairs to the cataclysmic variables 
suggest much longer, $\sim10^4$~years timescales for the CEE in these systems, possibly suggesting non-dynamical CEE in this regime. 
\end{abstract}




\section{Introduction}

The common envelope (CE) stage is an important stage in binary evolution, occurring in close binaries, typically when the primary evolves off the main sequence and expands. It gives rise to short period binaries, and drives the mergers of stars either directly or through the later evolution of post-CE compact remnant binaries that merge through gravitational-wave emission.  For a CE to ensue, the envelope needs to overflow the Roche lobe as to initiate a mass transfer to the companion. If and when the mass transfer is unstable, the primary's envelope engulfs the binary companion as to give rise to a CE. 
The following dynamical evolution is then driven by the gas drag force and gravity \citep{paczynskii1976} leading to the inspiral of the binary. The inspiral stage results in the binary merger or in the formation of a tight post-common envelope (pCE) binary (for a most recent review see \citealt{ivanova_cee_review}).

Purely hydrodynamical simulations do not give rise to the full ejection of the CE, in contrast with the observations of naked post-CE binaries \citep{ivanova_cee_review}. Such difficulties lead to the introduction of possible additional processes that may play a role in the CEE. These include the effects of recombination, dust driven winds or jets (see \citealt{2018MNRAS.478L..12G} for a brief overview). 

Here we consider several types of systems which likely went through a CE stage. In the case when the primary is low-mass star, the naked helium core is seen for a short time as subdwarf B stars (sdBs) before it turns into a white dwarf (WD), and therefore sdB stars are likely result of a post-CE evolution, a remnant post-CE WD could undergo a mass transfer from a close secondary star, in which case it is likely to manifest itself as a cataclysmic variable (CV, for review see \citealt{2010MmSAI..81..849R}). Finally, a non-accreting short period WD binary, is another possible post-CE product. 



The mass ejection during the CEE is typically thought to occur at relatively short (dynamical) timescales comparable to the inspiral time of the secondary or somewhat longer. If only a part of the envelope mass is ejected at the dynamical timescale, the CE could be initiated multiple times and, therefore, the mass loss timescale becomes longer. 
However, the timescale of CE, the minimum companion mass and the fraction of pre-CE to post-CE binary separations are not yet constrained observationally. Recently, \cite{2019MNRAS.484.4711M} tried to probe the timescale using two pCE binaries with additional wide astrometric components, i.e. using wide triple systems. We aim at extending this analysis and search for common proper motion and parallax pairs to pCE binaries, sdBs and CVs using a large sample based on the Gaia second data release \citep{2016A&A...595A...1G, 2018A&A...616A...1G}.



When the inner binary in a hierarchical triple enters the CE stage, the mass ejection strongly affects orbit of the distant third companion. 
We expect that the distribution of projected separations for ultra-wide components and fraction of survived ultra-wide components will therefore differ between the case of binaries which did not go through a CEE and ones which did lose mass through the CEE process.

Approximately 10 per cent of solar mass stars are born as hierarchical triples \citep{2017ApJS..230...15M}. This fraction reaches up to 20 per cent for stars extending to four solar masses. Triples are also known among systems which went through the CE evolution such as Wolf 1130 \citep{2013ApJ...777...36M} and GD 319 \citep{2005ApJS..161..394F}. These third components have orbital separations of $\approx 3.2\times 10^3$~a.u. in the case of Wolf 1130 and $\approx 5.5\times 10^4$~a.u. in the case of GD 319. Triples are also found among the sdB stars; for example, PG 1253+284 is seen as resolved pair with a separation of 0.24~arcsec and additionally shows radial velocity variations \citep{2002A&A...383..938H}.  Another possible case is SDSS J095101.28+034757.0 \citep{2015A&A...576A..44K} which shows an excess of IR radiation.

In the following we explore wide companions statistics using large samples. Our analysis follows similar ideas used in the work of \cite{2018MNRAS.480.4884E} who studied the projected separation of ultra-wide companions to WDs in order to infer natal kick magnitudes of WDs. Such analysis can provide the first statistical constraints on the CEE mass-loss time scales based on large samples. 

The paper is structured as follows: in Section~\ref{s:scenarios} we summarize the formation scenarios for systems which experience CEE. In Section~\ref{s:data} we describe our data set and in Section~\ref{s:method} we describe our method to search for ultra-wide binaries using the second Gaia data release. In Section~\ref{s:results_uwb} we describe the results of our search for ultra-wide components.
In Section~\ref{s:sim} we perform a simple simulation for orbital evolution of triple systems with a significant mass loss from the inner binary and we conclude with results and discussions in Section~\ref{s:res}.

\section{Formation paths of post common envelope binaries}
\label{s:scenarios}
In this Section we briefly describe scenarios suggested to explain the formation of post-common envelope binaries with a white dwarf and sdB stars.

\subsection{Formation of post-common envelope binaries with a white-dwarf component}
Low-mass white dwarfs with mass less than $\approx 0.5~M_\odot$ should  form through isolated stellar evolution only on timescales which greatly exceed the Hubble timescale. 
Nevertheless,  low-mass WDs are not rare among main sequence -- white dwarf binaries (MSWD) and contain up to a third of the observed population (see \citealt{2016MNRAS.458.3808R}). The MSWD sample is not complete and the actual fraction might differ due to various selection effects. The most natural explanation for their formation is through a CEE in binary systems when the more massive primary star losses its extended hydrogen envelope at the subgiant stage due to interaction of the envelope with the secondary main sequence  star, leaving behind the He core which later becomes a low-mass, typically He-rich WD (e.g. \cite{Zen+19} and references therein). Such binaries are seen as composite spectra binaries with large, periodic radial velocity variations observed through their optical spectra or as eclipsing binaries with reflection effects seen in the light-curve, e.g. HW Vir type systems \citep{2016PASP..128h2001H}. 


\subsection{Formation of sdB stars}
sdB stars are low-mass stars ($M\approx 0.5M_\odot$ \citealt{1986A&A...155...33H,2016PASP..128h2001H}) located to the left of the main sequence at the Hertzsprung Russell diagram around absolute magnitude of $\approx 5$. Most of these stars are considered to be helium-core burning stars with thin hydrogen envelopes which contain $<0.02~M_\odot$ \citep{1994ApJ...432..351S}. 
Such sdB stars experience a significant mass loss and many of them are found in close binaries with orbital periods of less than 10 days, which suggests a formation through a CE stage.

Following classical stellar evolution theory, an isolated red giant is not expected to lose its envelope and turn into a helium burning core. Therefore, the theories for the sdB formation include either non-standard stellar evolution (helium mixing, hot-flash) or the presence of the secondary companion which serves to strip the sdB stellar progenitors. 
When a sdB star is observed to be part of a close binary, the secondary star had to play an important role in the sdB formation.  Indeed, recent radial velocity measurements \citep{2004Ap&SS.291..321N,2011MNRAS.415.1381C} discovered a large binarity fraction among sdB stars with up 50 per cent or higher. In most cases the binary companions can not be directly detected, consistent with most of them being WDs or M-dwarfs.

Several binary evolution scenarios were suggested for the origin of sdB stars \citep{2002MNRAS.336..449H}. These include CE evolution \citep{paczynskii1976}, leading to the formation of very short period binaries with orbital periods of  less than 10 days; Roche-lobe overflow leading to the formation of wider binaries and mergers of two helium WDs \citep{1984ApJ...277..355W}, leading to the formation of single/isolated sdB stars. It is thought that CEE plays a key role in most cases and that up to 2/3 of known sdBs \citep{2002MNRAS.336..449H} are formed through this process. The expected mass of sdB progenitors \citep{2003MNRAS.341..669H} range between 0.9~$M_\odot$ (the lightest star which could form a red giant on timescales smaller than the Hubble time) to $\approx 3~M_\odot$. sdBs formed from a more massive progenitors are expected to be rare because of the initial mass function, shorter lifetime of massive He stars and other selection effects. Moreover systems with more massive progenitors are classified as sdO and Wolf-Rayet stars \citep{2019arXiv190806102G}.

\subsection{Cataclysmic variables}
A typical cataclysmic variable contains a CO WD component with a mass of $\approx 1\, M_\odot$ and a secondary with a mass of $\approx 1\, M_\odot$ \citep{2010MmSAI..81..849R}. The CO WD is, therefore, thought to originate from a primary in the mass range $2.2-8\, M_\odot$ because these WDs are formed following the complete loss of the hydrogen rich envelope at the AGB stage before the carbon burning has started \citep{2010MmSAI..81..849R}.  A CV might alternatively contain an ONe WD in which case it might have formed from even more massive stars.

The CE is initiated when the primary expands during its post-MS evolution. After the end of the CE, the semi-major axis of the binary shrinks following the loss of angular momentum through magnetic braking and/or gravitational wave emission. At some point the secondary fills its Roche lobe and a second mass transfer epoch is initiated. This second mass transfer is usually stable and the binary is seen as CV at this stage.

\section{Data}
\label{s:data}
In the following we describe the data collected for the various type of post-CE objects discussed above.
\subsection{Main sequence -- white dwarf binaries}
We use the catalogue of MSWD binaries compiled by \cite{2007MNRAS.382.1377R,2012MNRAS.419..806R,2013MNRAS.433.3398R,2016MNRAS.458.3808R}\footnote{https://www.sdss-wdms.org}. This catalogue includes 3287 MSWD binaries identified in the Sloan Digital Sky Survey \citep{sdss}. Only $\approx 25$ per cent of MSWD binaries are classified as post-CE systems. This small fraction is a consequence of the technique applied:  \cite{2007MNRAS.382.1377R} considered that an MSWD is a possible post-CE system only if it was observed during multiple epochs and it showed radial velocity variations. Meanwhile, among $\approx 3000$ MSWD systems only $\approx 600$ were observed during multiple epochs. In order to get the parallaxes and proper motions for these stars we match these data with the second Gaia data release \citep{2016A&A...595A...1G, 2018A&A...616A...1G}. All details of the cross match are summarized in Appendix~\ref{s:identific}.

\subsection{Hot subdwarf systems}
We use the catalogue of sdBs by \cite{2019A&A...621A..38G}. This catalogue contains 39800 candidates selected in Gaia DR2 and includes some possible contamination at the level of 10~per~cent. Only 9826 objects from this catalogue satisfy our quality cuts (where we require the relative errors in the parallax and in the proper motion measurements to be below 0.25). The measured parallaxes range from 0.1~mas up to 56~mas. In order to be sensitive to ultra-wide binaries we select only systems with measured parallaxes larger than 0.66~mas. Also we exclude two systems: HD110698 and BD+164120B for which we had troubles accessing the GAIA database. After these additional cuts we end up with 4709 systems.

We also tried to consider the recent catalogue by \cite{2019MNRAS.486.2169K}. Unfortunately, we manage to identify only 259 sdBs stars from this catalogue in the Gaia database using SDSS i,g colours with the conversion by \cite{jordi2010} within 3~arcsec from their catalogue positions. From this list of 259 stars only 69 have well measured parallax and proper motions which are suitable for our ultra-wide binary search, but we found no ultra-wide binary counterparts for any of them.

\subsection{Cataclysmic variables}
As a source for positions of cataclysmic variables we used the catalogue of \cite{2003A&A...404..301R} v.7.20\footnote{ http://www.MPA-Garching.MPG.DE/RKcat/ } which contains 1429 objects.  Because CVs are very variable in the optical band, we used only the coordinate information in our identification and did not perform magnitude or color analysis.
We searched for Gaia counterparts within 1.8~arcsec of the given catalogue positions and managed to identify 562 objects with good astrometric measurements of the parallax and the proper motion. 

We also notice that some of sdB objects listed in catalogue by \cite{2019A&A...621A..38G} are in fact CVs and we exclude them from our analysis of sdBs.

\subsection{Comparison samples}
Hierarchical triple systems with central pCE or MSWD wide binary  originate from hierarchical main sequence systems.  
Therefore, we want to identify wide binaries to main sequence stars and compare their occurrence rates with wide binaries to pCE systems. We select three samples: (A) direct comparison to sdBs, (B) a sample of only close-by objects (parallax $\varpi > 5$~mas) and (C) more massive stars to be compared with CVs (which likely originate from more massive stars). The ADQL requests are summarized in Appendix~\ref{a:adql_compar}.

Comparison sample A contains 10000 main sequence stars selected by the stellar radius and temperatures determined by the classification algorithm Apsis \citep{2013A&A...559A..74B,2018A&A...616A...8A}. We chose stars more massive than the Sun with masses $2- 3~M_\odot$ and with relative errors in parallax and proper motion measurements of less than 0.2. We also restrict the measured parallax to be in the range 0.67-10~mas as to select this sample in exactly the way we have selected the sdBs.

Comparison sample B contains 2452 stars with parallax $\varpi > 5$ and a relative error in parallax of less than 0.05. These stars are selected based on their color and absolute magnitude which are not corrected for extinction. We could not use the results of the Apsis algorithm for this sample because only a small number of stars were successfully classified using it. This sample is selected in such a way as to resolve ultra-wide binaries with separations of $10^2-10^3$~a.u. These binaries are the type of possible progenitors for sdBs with tertiary ultra-wide components at projected separations of few$\times 10^2-10^4$~a.u.

Comparison sample C contains 3399 stars. These are more massive stars (a minimal mass of 3.5~$M_\odot$, with a mean mass of $\approx 6\, M_\odot$, and a maximum mass of $9\, M_\odot$). Given the stellar initial mass function such stellar population is inherently less frequent and we therefore extended our selection up to parallax $\varpi>2$~mas in order to be able to identify sufficient number of appropriate stars and be able to resolve pairs with projected separations of $\sim 10^2$~a.u. In order to select stars for this sample, we require the relative error in the parallax and the proper motion to be less than 0.1. 
The sample is used to simulate the survival fractions of CVs (which typically originate from these more massive stars) with ultra-wide companions.

\section{Method}
\label{s:method}
We identify common proper motion and parallax pairs to MSWD, sdBs and CVs stars following the method described in \cite{2018MNRAS.480.4884E} and in our recent work  \citep{igoshev_massive}. 
We assume that two stars are likely to be gravitationally bound if they are located close at the sky, have similar parallaxes and move in similar directions.
Practically, we check if following criteria are satisfied: (1) their parallaxes differ by less than twice the error in the parallax difference; (2) the proper motion difference is less than twice the error in the proper motion difference plus the contribution due to the orbital motion; (3) the error in the parallax difference is below 0.6~mas; (4) the error in the proper motion difference is below two times the possible difference due to the orbital motion; and (5) the error in the proper motion difference is below 1.2~mas~year$^{-1}$.

For each of the cases where good astrometric quality was attained for our MSWD, sdB or CV systems, we selected all stars with projected spatial separations less than $2\times 10^5$~a.u. from the system, and identified potential companions  with good astrometric solution and relative errors in parallax, and proper motions smaller than 0.33 of their value from the second Gaia data release. Following this step, we then considered whether these potential targets met the five criteria mentioned above.

We also made an additional check, as to reject a possible spurious origin of a wide companion due to association with a cluster. In particular, we searched for any known open clusters in the catalogue by \cite{2018A&A...618A..93C} at angular separation of 1 degree with mean parallax difference of  less than 0.3~mas and a mean proper motion difference of  less than 2~mas~year$^{-1}$. Since this catalogue has no information about globular clusters, we also checked possible association to known globular clusters.

\section{Results}
\label{s:results_uwb}
\subsection{Comparison samples}
In comparison sample A we identify 498 wide companions, from which we infer a multiplicity fraction of $5.0\pm 0.2$ per cent, see Table~\ref{t:fract}. 
In order to estimate the contamination level, we then searched the positions of the same stars in the comparison sample but after shifting the locations by 1.0 degree in the declination direction (the largest size of the searching area for the sample), and performed the search again on this synthetic sample. In this case we find 103 ultra-wide companions for the stars in the synthetic shifted sample (where 9 belong to the open cluster NGC 2632, which we therefore excluded). From these results we infer a chance alignment of companion stars at the level of $0.94\pm 0.1$ per cent in our comparison sample.

\begin{figure}
\includegraphics[width=\columnwidth]{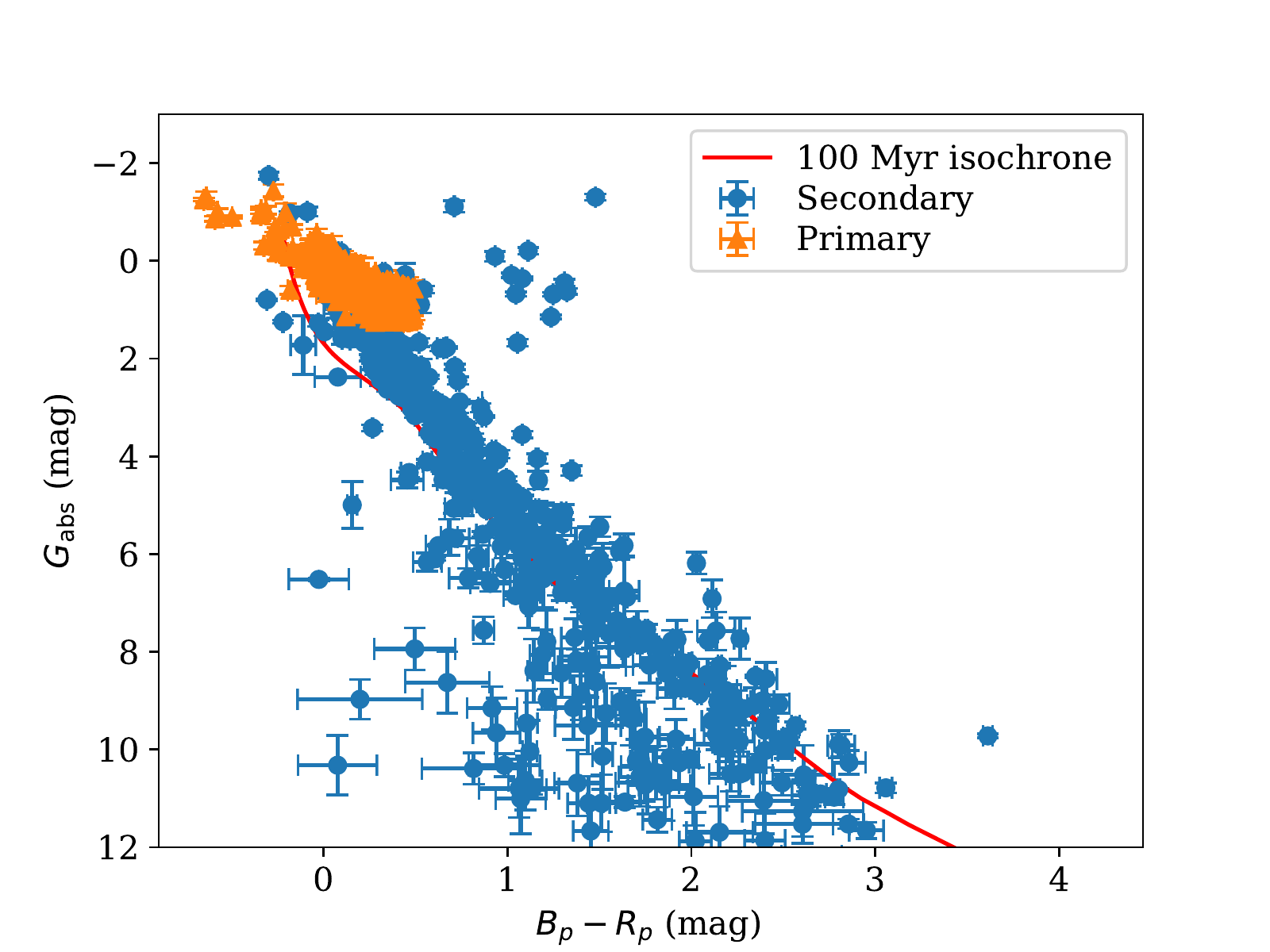}
\caption{The Hertzsprung Russel diagram for ultra-wide binaries found in comparison sample A.}
    \label{f:hr_compar}
\end{figure}

In Figure~\ref{f:hr_compar} we show the Hertzsprung Russel diagram for ultra-wide components. The color and magnitude data in this plot were corrected for reddening using the three-dimensional map of \cite{green_map}. For the conversion of $E(B-V)$ to $A_g$ and $E(B_p-R_p)$ we use fixed values of $R = 3.1$, $A_g/A_v = 0.9$ and $E(B_p-R_p)/E(B-V)=1.5$, and apply a factor of 0.884 to all the reddening values.

\begin{table}
    \centering
    \begin{tabular}{llccc}
    \hline
    Type             &  Ultra-wide   & Uncertainty     \\
                     & multiplicity fraction & \\
    \hline
     MS + distant (A)  &  $498/9934\approx 0.050$    &  0.002 \\
     MS + distant (B)  &  $197/2201\approx 0.089$    & 0.006 \\
     $M_\mathrm{third} > 0.4\, M_\odot$ \\
     MS + distant (B)  &  $155/2201\approx 0.070 $   & 0.005\\
     $M_\mathrm{third} > 0.6\, M_\odot$ \\
     MS + distant (C)  & $161/3399\approx 0.047$     & 0.004 \\ 
     sdB + distant     &  $68/4709\approx 0.014 $    & 0.002\\
     MSWD + distant    &  $42/998\approx 0.042 $     & 0.006 \\
     no CEE \\
     pCE + distant     &  $6/161\approx 0.037 $      & 0.015  \\
     CVs + distant     &  $14/562\approx 0.025 $     & 0.007\\
     \hline
    \end{tabular}
    \caption{Ultra-wide binarity/multiplicity fraction found in our research. }
    \label{t:fract}
\end{table}

\begin{figure}
\includegraphics[width=\columnwidth]{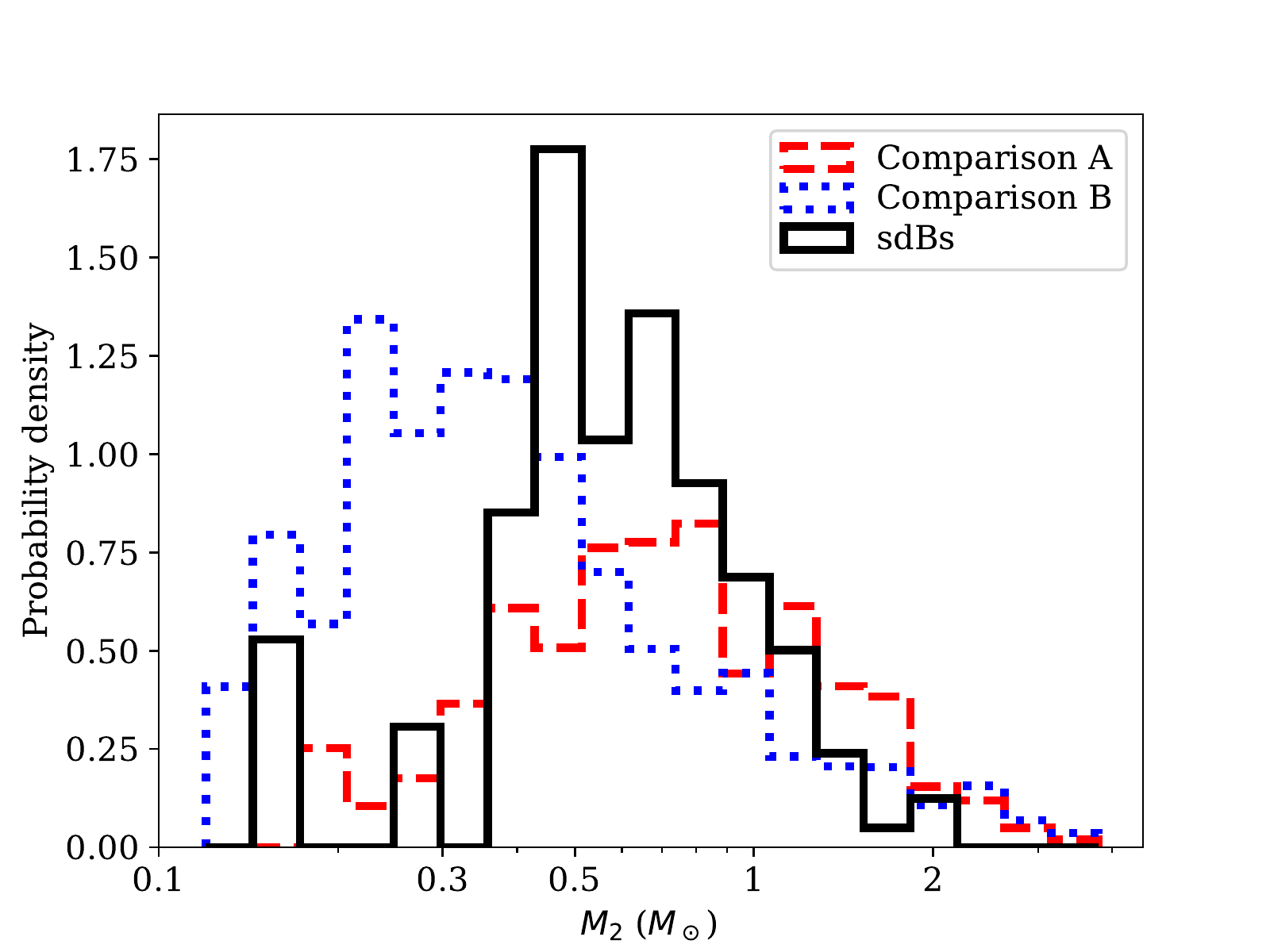}
\caption{The distribution of masses of ultra-wide companions for sdB stars and the comparison samples A and B.}
    \label{f:sdb_secondary_mass}
\end{figure}

It seems that the stellar parameters determined by the Apsis algorithm indeed place primary stars at the main sequence just above the Sun and below an absolute magnitude of $G_\mathrm{abs}=0$. The secondary stars are mostly low-mass stars with a mass distribution peaking at $M = 0.7~M_\odot$, see in  Figure~\ref{f:sdb_secondary_mass}. 

\begin{figure}
\includegraphics[width=\columnwidth]{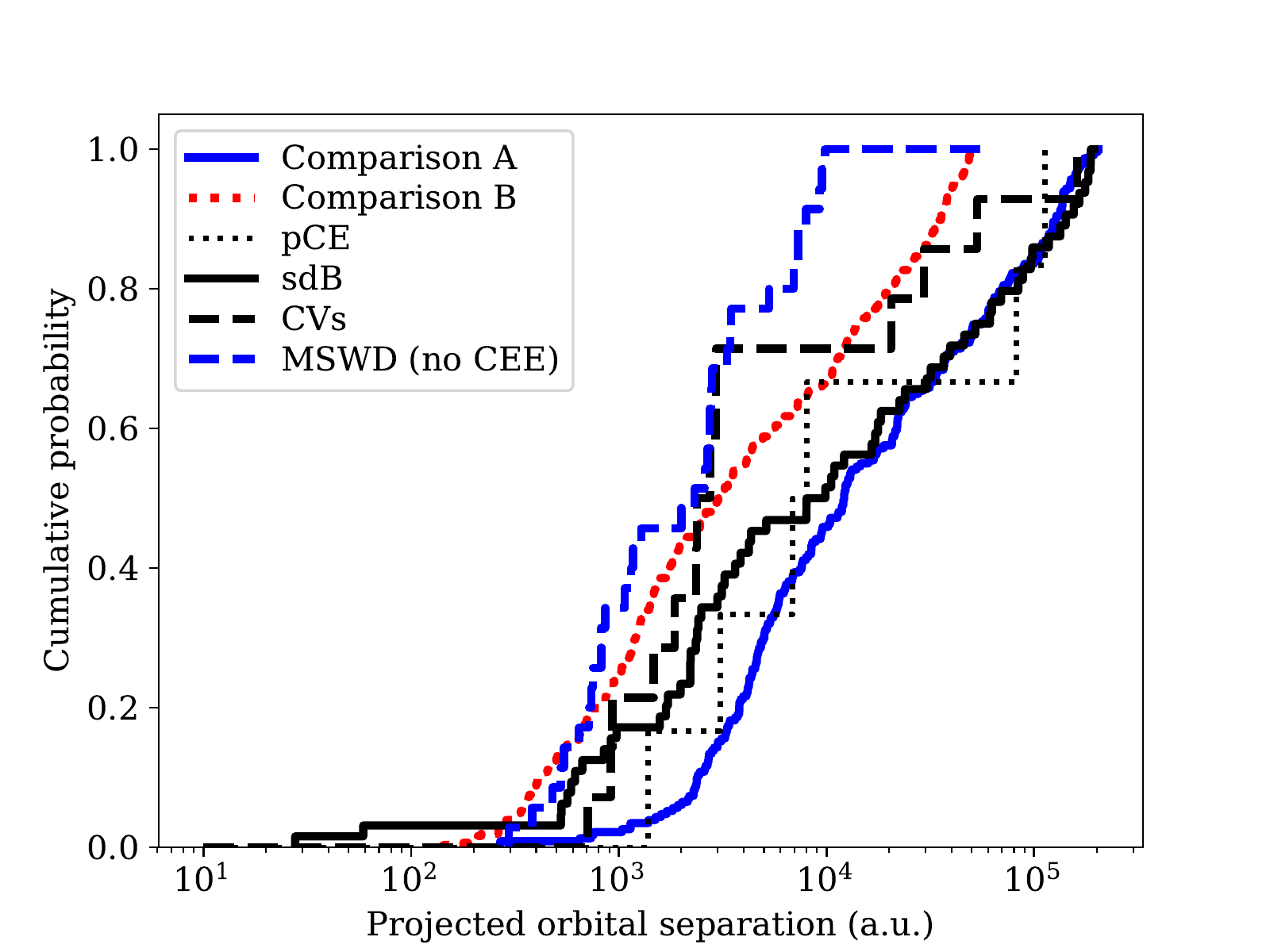}
\caption{The cumulative distribution of orbital separations for the various post-CE systems and the comparison samples. Shown are ultra-wide companions to sdB stars (solid black line), MSWD stars (dashed blue line; excluding close resolved binaries and ones with CEE) and short-period pCE-binaries (dotted black line), CVs (dashed black line) and the comparison samples A and B (excluding secondaries with masses less than $0.4\, M_\odot$). }
    \label{f:cum_orb_sep}   
\end{figure}

The distribution of projected separations is shown in Figure~\ref{f:cum_orb_sep}. In comparison to the work by \cite{2018MNRAS.480.4884E} we extend the radius of the searching region up to $2\times 10^5$~a.u. and estimate the total ultra-wide binarity fraction. It is also worth noting the following effect: in the Gaia DR2 two stars are considered as separate stars with reliable photometry if the angular separation between them is more than 0.5-1~arcsec. It means that in the \cite{2018MNRAS.480.4884E} sample the resolution is always better than 200~a.u. In our sample A most of the stars are located at typical distances of 1~kpc, and therefore the typical resolution is of the order of 1000~a.u. 

In comparison sample B we initially identify 323 ultra-wide binaries. We then excluded all systems with more than one ultra-wide components as to get better resemblance to work by \cite{2018MNRAS.480.4884E}. 
Since this (B) comparison sample is of stars located much closer to us, it enables a better sensitivity to much fainter secondaries than in the sdBs sample (see Figure~\ref{f:sdb_secondary_mass}). Therefore, we considered two additional cuts on the companion mass, in order to enable a proper comparison of the different sample. In particular, in one case we excluded all the systems where the secondary mass was less than $0.4\, M_\odot$ and in the second we considered a mass cut-off of $0.6\, M_\odot$. The results are summarized in the Table~\ref{t:fract}. The distribution of the projected separations for sample B with the  $0.4\, M_\odot$ cut-off is shown in Figure~\ref{f:cum_orb_sep}.

\begin{figure}
\includegraphics[width=\columnwidth]{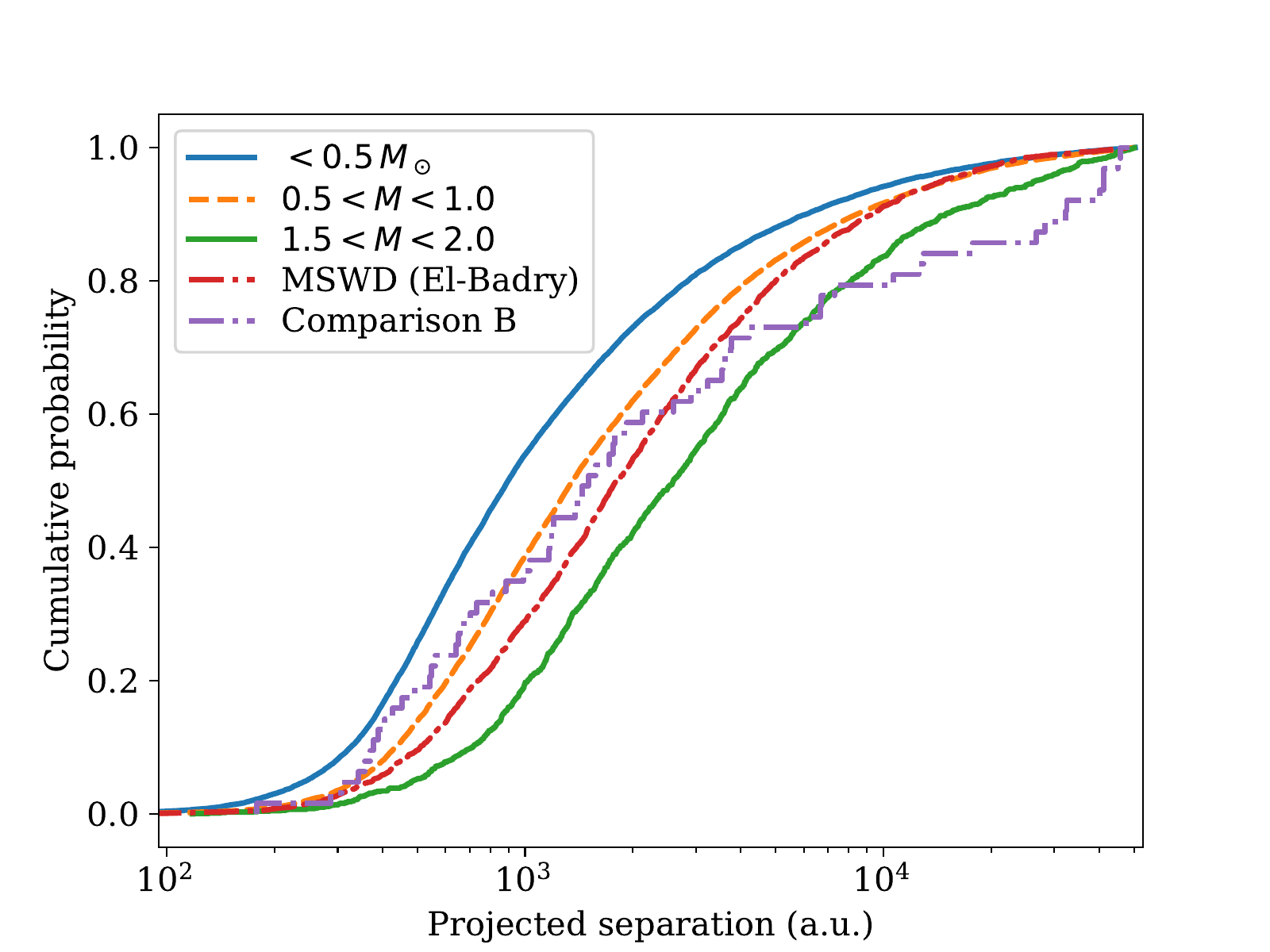}
\caption{The cumulative distribution of projected separations to ultra-wide companion (with a cut at $<5\times 10^4$~a.u.) for the sample of  \protect\cite{2018MNRAS.480.4884E} and for the more massive stars from our comparison sample (this work).}
    \label{f:mass_a}
\end{figure}

Besides a shift in the cumulative distribution which could be caused by our limited resolution in comparison to \cite{2018MNRAS.480.4884E}, we see a clear trend for increasing projected separation of the ultra-wide companion with increasing mass of the primary star, see Figure~\ref{f:mass_a}\footnote{This effect could be a consequence of the observational selection. More massive primaries tend to be further away and given a cut in distances they are more rare in the sample, so the sample becomes incomplete for projected separations of a few hundred AU.}. We estimated the masses of the stars in the \cite{2018MNRAS.480.4884E} sample using a combined isochrone (an age of 2~Myr for $M>3.5\, M_\odot$, an age of  10~Myr for $1.8<M<3.5\, M_\odot$ and $M<1.8\, M_\odot$ and an age of 0.5~Gyr\footnote{These ages are unimportant for the following analysis since we are mostly interested in the stellar mass at the zero age mass sequence and mass loss is quite small for low-mass stars.}). When possible we corrected for absorption using  \cite{green_map}. We find the difference between the cumulative distribution of MSWD ultra-wide binaries and that of MSMS wide binaries to be smaller in comparison with the difference between the cumulative distribution of the projected separations for ultra-wide components when the primary mass is less than $0.5~M_\odot$ and the primary mass is in the range $1.5<M<2\, M_\odot$.

In the comparison sample C (massive primaries) we identify 297 ultra-wide pairs. After we exclude repetitions and secondaries with masses less than $0.4\, M_\odot$ we are left with 180 objects. This results in a multiplicity fraction of $4.7\pm 0.4$~per~cent.


\subsection{MSWD and PCE systems}
\begin{figure*}
\begin{minipage}{0.48\linewidth}
	\includegraphics[width=\columnwidth]{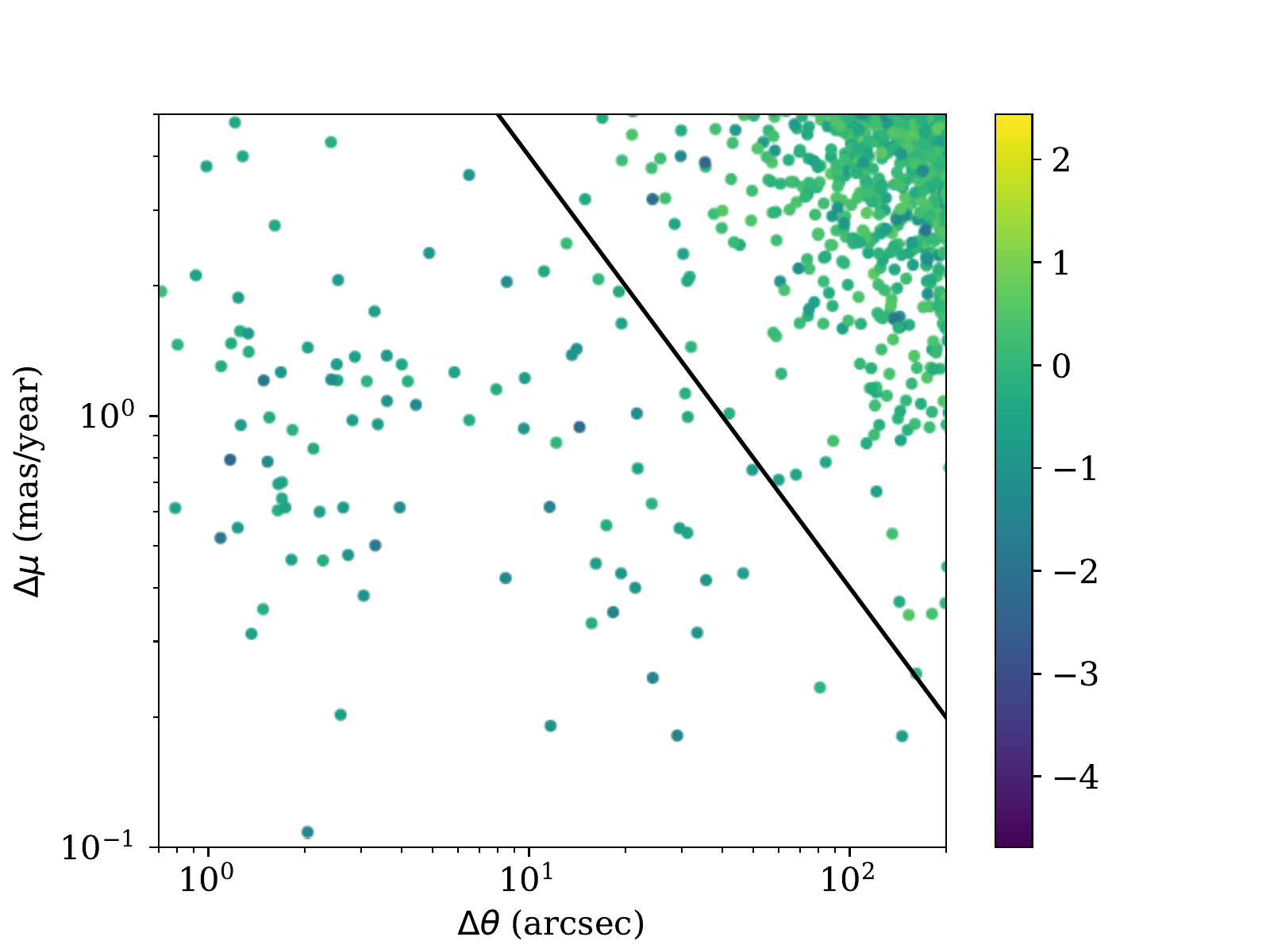}
\end{minipage}
\begin{minipage}{0.48\linewidth}
	\includegraphics[width=\columnwidth]{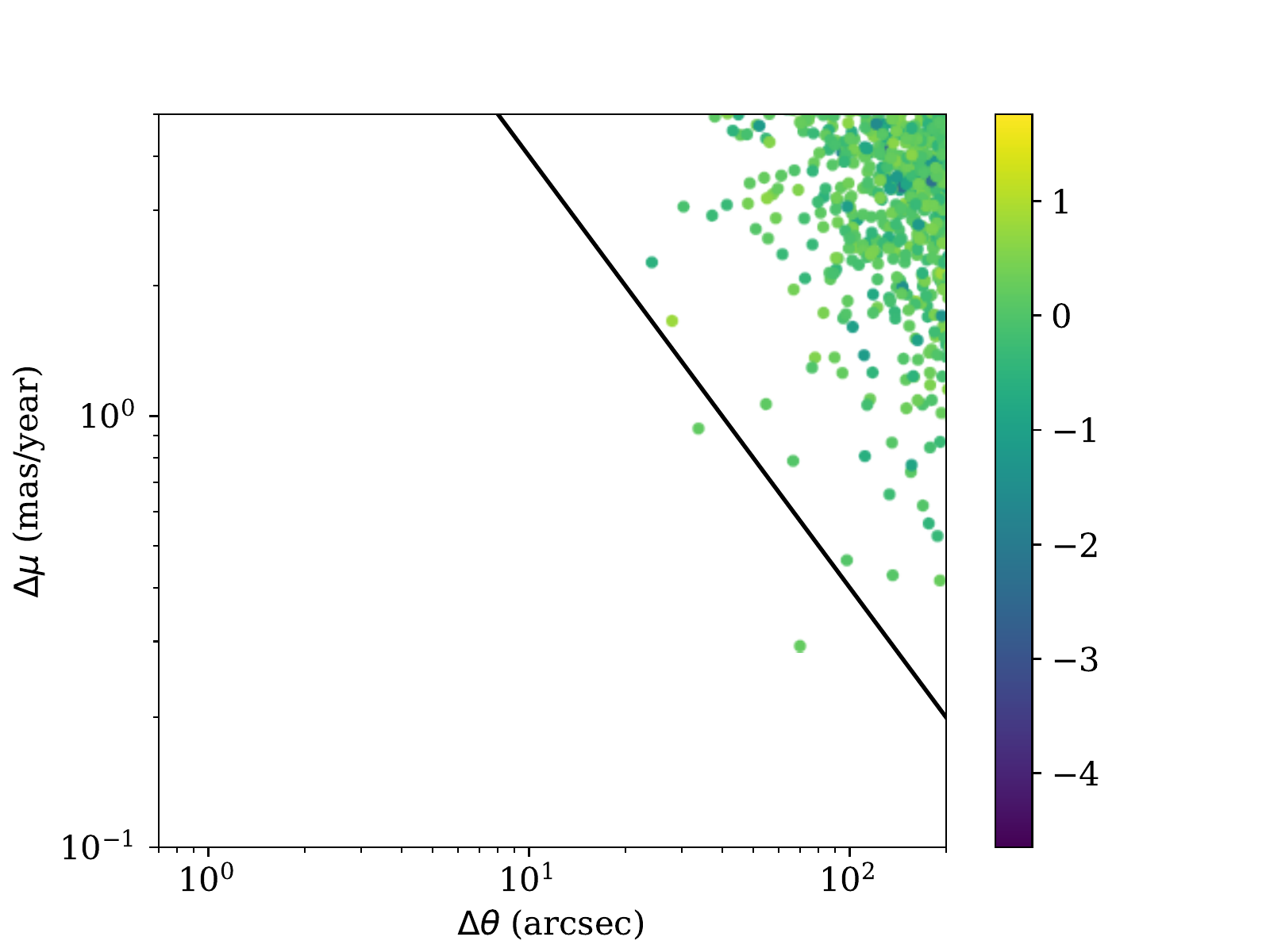}
\end{minipage}
\caption{The difference in proper motions vs. angular separations for MSWD systems. Left panel: the real  sample. Right panel: the results for the mock, shifted sample (1.4 degrees in the direction of right ascension). The color shows the logarithm of the parallax difference. The black solid line guides the comparison between the real and the shifted samples.}
    \label{f:nswd_ang}
\end{figure*}

In our analysis we have identified 63 common proper motions and parallax pairs to MSWD systems (see the distribution of the differences in position and proper motion in Figure~\ref{f:nswd_ang}). It is worth noting that the SDSS spectroscope uses fibers with a diameter of 3~arcsec on the sky \citep{sdss}. Thereby, there is a number of binaries which are resolved in the Gaia database (especially by the astro broad band photometer with the angular resolution up to 0.1~arcsec), but are considered to be spectroscopic binaries in our MSWD sample. To deal with this problem we further divide our list into two parts. The common proper motion and parallax pairs with angular separations of less than 2~arcsec (21 pair) are considered to be \textit{resolved binaries}, see Table~\ref{t:resolved}, while the 42 pairs with angular separations larger than 2~arcsec are considered to be \textit{triples with ultra-wide companion}, see Table~\ref{t:b_candidate}. This division has a certain degree of arbitrariness, but it is impossible to make a better choice without additional observations. The third Gaia data release will provide information about the radial velocities and the binary properties which will help to better separate the samples.

We plot the projected orbital separations in Figure~\ref{f:cum_orb_sep}.  Six binaries which are marked as the post-common envelope systems in the catalogue seems to follow a much wider projected separation distribution than the general MSWD binaries which did not go through a CE episode. In particular, two systems with separations larger than $8\times 10^4$~a.u. are pCE binaries with third distant components.


\begin{figure}
	\includegraphics[width=\columnwidth]{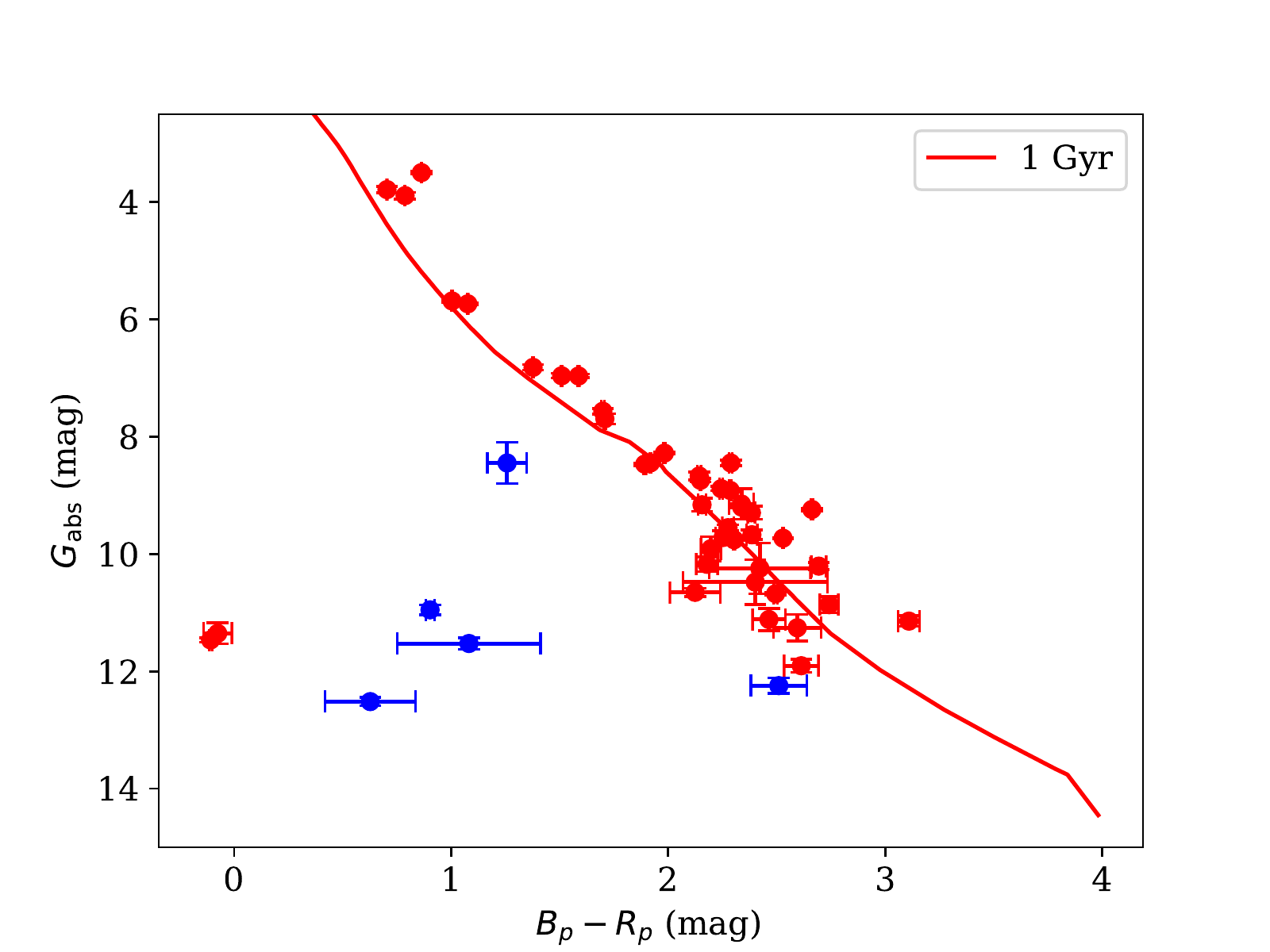}
\caption{The Hertzsprung Russell diagram for the third (red dots) and the second (blue dots) components of MSWD systems.
}
\label{f:hr_mass}
\end{figure}

We also plot the Hertzsprung Russel diagram for ultra-wide components, see Figure~\ref{f:hr_mass}.
We see that the majority of the ultra-wide pairs are low-mass main sequence stars with spectral types G or K. Five objects are WDs, most of which are the components of the resolved binaries,  supporting our original sample division. Two white dwarfs with color $B_p - R_p \approx -0.1$ are well separated from the MSWD binaries (11 and 50 arcsec) which means that they are actual tertiary companions and not resolved MSWDs. For a large number of resolved binaries Gaia colors are not provided because the angular resolution of the medium-band photometers of the Gaia is 0.5-1~arcsec \citep{2006MNRAS.367..290J} and the components of the binary are not resolved.

\begin{table*}
    \centering
    \begin{tabular}{llccrcccrcccc}
        \hline
        Name        &  Gaia primary & Gaia secondary & $\Delta \theta$  & $\Delta\varpi \pm \sigma_{\Delta\varpi}$ & $\Delta \mu\pm \sigma_{\Delta\mu}$ & $A$  & pCE \\
                    & Gaia DR2 name & Gaia DR2 name  & (arcsec)           & (mas)                                          & (mas year$^{-1}$)                        & (a.u.)\\
        \hline
SDSSJ011055.30-102011.9  &  2469937118135492480  &  2469937118136325632  &  1.7  & $ 0.188  \pm  0.253 $ & $ 0.692  \pm  0.381  $ &  234.1  &  N  \\
SDSSJ024519.11+011157.3  &  2499299159543260928  &  2499299159545173760  &  1.7  & $ 0.505  \pm  0.5 $ & $ 0.601  \pm  0.734  $ &  761.1  &  N  \\
SDSSJ025202.46-010515.7  &  2497494654803989760  &  2497494654805378816  &  1.7  & $ 0.303  \pm  0.179 $ & $ 0.611  \pm  0.276  $ &  985.1  &  N  \\
SDSSJ081327.92+373245.6  &  907874108333645312  &  907874108334409856  &  0.8  & $ 0.789  \pm  0.544 $ & $ 1.454  \pm  1.177  $ &  472.9  &  N  \\
SDSSJ084518.66+055911.7  &  583017522392120064  &  583017522392559232  &  1.3  & $ 0.368  \pm  0.395 $ & $ 3.966  \pm  1.186  $ &  352.2  &  N  \\
SDSSJ091508.22+415559.5  &  816062001197670016  &  816062001196480256  &  1.7  & $ 0.114  \pm  0.18 $ & $ 1.256  \pm  0.241  $ &  441.8  &  N  \\
SDSSJ092203.36+394002.0  &  812448078274747520  &  812448078276868096  &  1.5  & $ 0.045  \pm  0.234 $ & $ 0.78  \pm  0.283  $ &  357.1  &  N  \\
SDSSJ103955.45+310643.5  &  736093077399500032  &  736093073107320192  &  1.7  & $ 0.381  \pm  0.559 $ & $ 0.641  \pm  0.716  $ &  225.7  &  N  \\
SDSSJ105845.26+164714.9  &  3981879852457277440  &  3981879852457628416  &  1.5  & $ 0.013  \pm  0.245 $ & $ 1.202  \pm  0.521  $ &  523.4  &  N  \\
SDSSJ111615.73+590509.3  &  857718476684324992  &  857718476683372288  &  1.1  & $ 0.015  \pm  0.39 $ & $ 0.519  \pm  0.857  $ &  428.4  &  N  \\
SDSSJ112118.04+585036.4  &  857547532690379008  &  857547536985870592  &  1.4  & $ 0.23  \pm  0.168 $ & $ 0.311  \pm  0.272  $ &  692.4  &  N  \\
SDSSJ114913.52-014728.6  &  3794340723954133504  &  3794340719659710720  &  1.7  & $ 0.307  \pm  0.289 $ & $ 0.699  \pm  0.367  $ &  336.2  &  N  \\
SDSSJ131156.69+544455.8  &  1564508327957813632  &  1564508327956853888  &  1.3  & $ 0.155  \pm  0.121 $ & $ 1.543  \pm  0.184  $ &  307.2  &  N  \\
SDSSJ134624.89+021734.2  &  3665130240625799808  &  3665130236330929792  &  1.2  & $ 0.181  \pm  0.195 $ & $ 1.867  \pm  0.333  $ &  248.9  &  N  \\
SDSSJ135907.48+294209.3  &  1453655286472125440  &  1453655286473655680  &  1.8  & $ 0.762  \pm  0.412 $ & $ 0.924  \pm  0.644  $ &  323.2  &  N  \\
SDSSJ152826.04+155916.4  &  1207541153468105344  &  1207541153466703872  &  1.5  & $ 0.569  \pm  0.302 $ & $ 0.355  \pm  0.398  $ &  791.4  &  N  \\
SDSSJ155232.50+202715.3  &  1204454485026467200  &  1204454485024334720  &  0.9  & $ 0.252  \pm  0.285 $ & $ 2.104  \pm  0.341  $ &  259.8  &  N  \\
SDSSJ155735.37+155817.2  &  1193136971322193792  &  1193136971325031296  &  1.2  & $ 0.004  \pm  0.447 $ & $ 0.788  \pm  0.516  $ &  695.7  &  N  \\
SDSSJ170127.36+253302.6  &  4573134327556676224  &  4573134323259272064  &  1.8  & $ 0.26  \pm  0.131 $ & $ 0.462  \pm  0.191  $ &  396.2  &  N  \\
SDSSJ172439.05+551600.1  &  1419718383339799168  &  1419718379044501632  &  1.2  & $ 0.18  \pm  0.283 $ & $ 0.548  \pm  0.592  $ &  746.7  &  N  \\
SDSSJ233919.64-000233.4  &  2642852260954014848  &  2642852260954740480  &  1.1  & $ 0.542  \pm  0.476 $ & $ 1.297  \pm  0.702  $ &  258.8  &  N  \\
    \hline
    \end{tabular}
    \caption{Resolved binaries found in the MSWD sample. $A$ is the projected orbital separation.}
    \label{t:resolved}
\end{table*}

\begin{table*}
    \centering
    \begin{tabular}{llccrcccrcccc}
        \hline
        Name        &  Gaia primary & Gaia secondary & $\Delta \theta$  & $\Delta\varpi \pm \sigma_{\Delta\varpi}$ & $\Delta \mu\pm \sigma_{\Delta\mu}$ & $A$  & pCE \\
                    & Gaia DR2 name & Gaia DR2 name  & (arcsec)           & (mas)                                          & (mas year$^{-1}$)                        & (a.u.)\\
        \hline
SDSSJ002428.44-102443.5  &  2424897475435562368  &  2424897479729961984  &  4.0  & $ 0.059  \pm  0.412 $ & $ 0.611  \pm  0.477  $ &  1281.2  &  N  \\
SDSSJ014246.00-094731.0  &  2464385576553809792  &  2464385580847869952  &  2.8  & $ 0.152  \pm  0.419 $ & $ 0.972  \pm  0.654  $ &  643.4  &  N  \\
SDSSJ022615.69-010423.9  &  2499922071535617280  &  2499922067240455168  &  6.5  & $ 0.501  \pm  0.357 $ & $ 0.973  \pm  0.553  $ &  2730.4  &  N  \\
SDSSJ023650.60-010313.3  &  2497111028324980736  &  2497111097044458880  &  21.9  & $ 0.257  \pm  0.302 $ & $ 0.752  \pm  0.451  $ &  9503.0  &  N  \\
SDSSJ024642.55+004137.2  &  2499031084864744448  &  2499031192239169280  &  18.3  & $ 0.029  \pm  0.195 $ & $ 0.35  \pm  0.317  $ &  6875.4  &  Y  \\
SDSSJ030607.18-003114.4  &  3266296412128039424  &  3266279026100283008  &  778.2  & $ 0.124  \pm  0.089 $ & $ 0.319  \pm  0.121  $ &  112642.6  &  Y  \\
SDSSJ030716.44+384822.8  &  142664833456549120  &  142664833456549376  &  3.6  & $ 0.085  \pm  0.195 $ & $ 1.077  \pm  0.22  $ &  1378.4  &  Y  \\
SDSSJ032510.84-011114.1  &  3262517837340737152  &  3262517841635204608  &  2.9  & $ 0.205  \pm  0.119 $ & $ 1.364  \pm  0.218  $ &  295.8  &  N  \\
SDSSJ080120.47+064614.7  &  3144220281799428736  &  3144220286094183680  &  3.3  & $ 0.013  \pm  0.19 $ & $ 0.499  \pm  0.2  $ &  822.6  &  N  \\
SDSSJ081647.38+534017.8  &  1031806794114311552  &  1031806798409670016  &  11.7  & $ 0.076  \pm  0.139 $ & $ 0.191  \pm  0.195  $ &  3328.7  &  N  \\
SDSSJ082823.55+470001.3  &  930577850922831616  &  930577855217008000  &  4.5  & $ 0.04  \pm  0.161 $ & $ 1.055  \pm  0.229  $ &  717.3  &  N  \\
SDSSJ085426.25+374653.0  &  719483236276555008  &  719483236276554880  &  2.6  & $ 0.164  \pm  0.173 $ & $ 0.611  \pm  0.234  $ &  479.0  &  N  \\
SDSSJ091218.46+150334.4  &  607478387644308736  &  607478391935326848  &  2.1  & $ 0.423  \pm  0.378 $ & $ 0.835  \pm  0.506  $ &  752.2  &  N  \\
SDSSJ093809.28+143037.0  &  617887567299257600  &  617887567299258752  &  11.2  & $ 0.369  \pm  0.393 $ & $ 2.149  \pm  0.457  $ &  2676.9  &  N  \\
SDSSJ095756.81+361444.9  &  796612911812751616  &  796612843095588736  &  2.0  & $ 0.226  \pm  0.435 $ & $ 1.431  \pm  0.524  $ &  382.1  &  N  \\
SDSSJ101958.61+283339.8  &  741061353833562880  &  741061353833562368  &  24.4  & $ 0.038  \pm  0.371 $ & $ 0.246  \pm  0.544  $ &  7299.7  &  N  \\
SDSSJ102118.15+265101.1  &  728746686163222272  &  728746686163197440  &  9.7  & $ 0.118  \pm  0.259 $ & $ 0.93  \pm  0.257  $ &  2315.9  &  N  \\
SDSSJ104959.80-004719.0  &  3803142859993965952  &  3803142829929703552  &  14.4  & $ 0.006  \pm  0.173 $ & $ 0.937  \pm  0.287  $ &  2783.4  &  N  \\
SDSSJ105607.54+583943.3  &  860485462121503360  &  860485466415271680  &  2.2  & $ 0.188  \pm  0.254 $ & $ 0.597  \pm  0.323  $ &  822.7  &  N  \\
SDSSJ105806.04+152225.9  &  3969333600151218176  &  3969333600151218304  &  8.5  & $ 0.04  \pm  0.12 $ & $ 0.419  \pm  0.182  $ &  5288.5  &  N  \\
SDSSJ111046.29+612225.2  &  861984959757517184  &  861984955461732992  &  29.1  & $ 0.027  \pm  0.223 $ & $ 0.181  \pm  0.38  $ &  9858.0  &  N  \\
SDSSJ114716.07+293930.3  &  4020741021494570624  &  4020741025789617536  &  31.3  & $ 0.286  \pm  0.297 $ & $ 0.533  \pm  0.45  $ &  9271.6  &  N  \\
SDSSJ115553.94+105255.2  &  3918510771102102528  &  3918510839821579392  &  21.5  & $ 0.109  \pm  0.166 $ & $ 0.398  \pm  0.164  $ &  7970.8  &  N  \\
SDSSJ115848.87+171553.1  &  3926599225312457472  &  3926599122233242112  &  33.5  & $ 0.087  \pm  0.226 $ & $ 0.313  \pm  0.348  $ &  7300.9  &  N  \\
SDSSJ124808.93+605726.4  &  1579901250228323584  &  1579901250228323712  &  2.6  & $ 0.195  \pm  0.123 $ & $ 0.202  \pm  0.172  $ &  738.2  &  N  \\
SDSSJ124959.75+035726.6  &  3705361680324471424  &  3705362504958192512  &  146.0  & $ 0.204  \pm  0.108 $ & $ 0.181  \pm  0.153  $ &  59698.3  &  N  \\
SDSSJ142149.14+382833.3  &  1484715149927519104  &  1484715154222882176  &  3.3  & $ 0.244  \pm  0.308 $ & $ 1.736  \pm  0.511  $ &  544.3  &  N  \\
SDSSJ142951.19+575949.0  &  1611769731470117120  &  1611769735766328576  &  16.2  & $ 0.161  \pm  0.16 $ & $ 0.453  \pm  0.276  $ &  8071.5  &  Y  \\
SDSSJ143642.01+574146.3  &  1611031615570789760  &  1611034334286174080  &  324.6  & $ 0.117  \pm  0.127 $ & $ 0.248  \pm  0.229  $ &  82181.7  &  Y  \\
SDSSJ145248.79+234807.6  &  1266149972245838720  &  1266149972245864576  &  5.9  & $ 0.213  \pm  0.292 $ & $ 1.255  \pm  0.52  $ &  2813.7  &  N  \\
SDSSJ145642.71+053101.8  &  1159963910942567168  &  1159963983957684736  &  21.7  & $ 0.052  \pm  0.18 $ & $ 1.008  \pm  0.327  $ &  3432.3  &  N  \\
SDSSJ153009.49+384439.8  &  1387898512537120768  &  1387898516831996544  &  4.0  & $ 0.28  \pm  0.192 $ & $ 1.309  \pm  0.311  $ &  2005.6  &  N  \\
SDSSJ154843.79+372749.7  &  1376105769292093184  &  1376105739228731392  &  19.4  & $ 0.148  \pm  0.136 $ & $ 0.43  \pm  0.274  $ &  3091.6  &  Y  \\
SDSSJ162020.89+214542.9  &  1298515020427786624  &  1298515024723278336  &  2.0  & $ 0.041  \pm  0.159 $ & $ 0.108  \pm  0.244  $ &  524.8  &  N  \\
SDSSJ170546.61+274028.3  &  4574942916809993088  &  4574942916808430336  &  2.5  & $ 0.187  \pm  0.393 $ & $ 1.31  \pm  0.599  $ &  860.9  &  N  \\
SDSSJ173430.11+335407.5  &  4602418200558831232  &  4602418204854379520  &  3.6  & $ 0.161  \pm  0.223 $ & $ 1.371  \pm  0.451  $ &  1069.7  &  N  \\
SDSSJ192306.01+620310.7  &  2240323111314621056  &  2240323111318292096  &  3.4  & $ 0.168  \pm  0.316 $ & $ 0.951  \pm  0.603  $ &  2602.3  &  N  \\
SDSSJ192616.13+383400.8  &  2052736600737294336  &  2052736600733320448  &  3.1  & $ 0.078  \pm  0.219 $ & $ 0.382  \pm  0.419  $ &  1144.8  &  N  \\
SDSSJ204713.67+002203.8  &  4228388774562523264  &  4228388602763827584  &  49.8  & $ 0.187  \pm  0.257 $ & $ 0.746  \pm  0.319  $ &  6967.9  &  N  \\
SDSSJ213225.96+001430.5  &  2687732916851442304  &  2687733015641916544  &  2.4  & $ 0.073  \pm  0.49 $ & $ 1.207  \pm  0.911  $ &  1170.5  &  N  \\
SDSSJ230202.49-000930.0  &  2651675425155232128  &  2651675051493595392  &  14.1  & $ 0.093  \pm  0.246 $ & $ 1.421  \pm  0.248  $ &  3479.1  &  N  \\
SDSSJ233919.64-000233.4  &  2642852260954014848  &  2642852329674217088  &  11.6  & $ 0.031  \pm  0.228 $ & $ 0.612  \pm  0.366  $ &  2733.4  &  N  \\
    \hline
    \end{tabular}
    \caption{Wide binaries found in the MSWD sample. $A$ is the projected orbital separation.}
    \label{t:b_candidate}
\end{table*}

\subsection{Hot subdwarf systems}
\label{s:sdb_res}
\begin{figure*}
\begin{minipage}{0.48\linewidth}
	\includegraphics[width=\columnwidth]{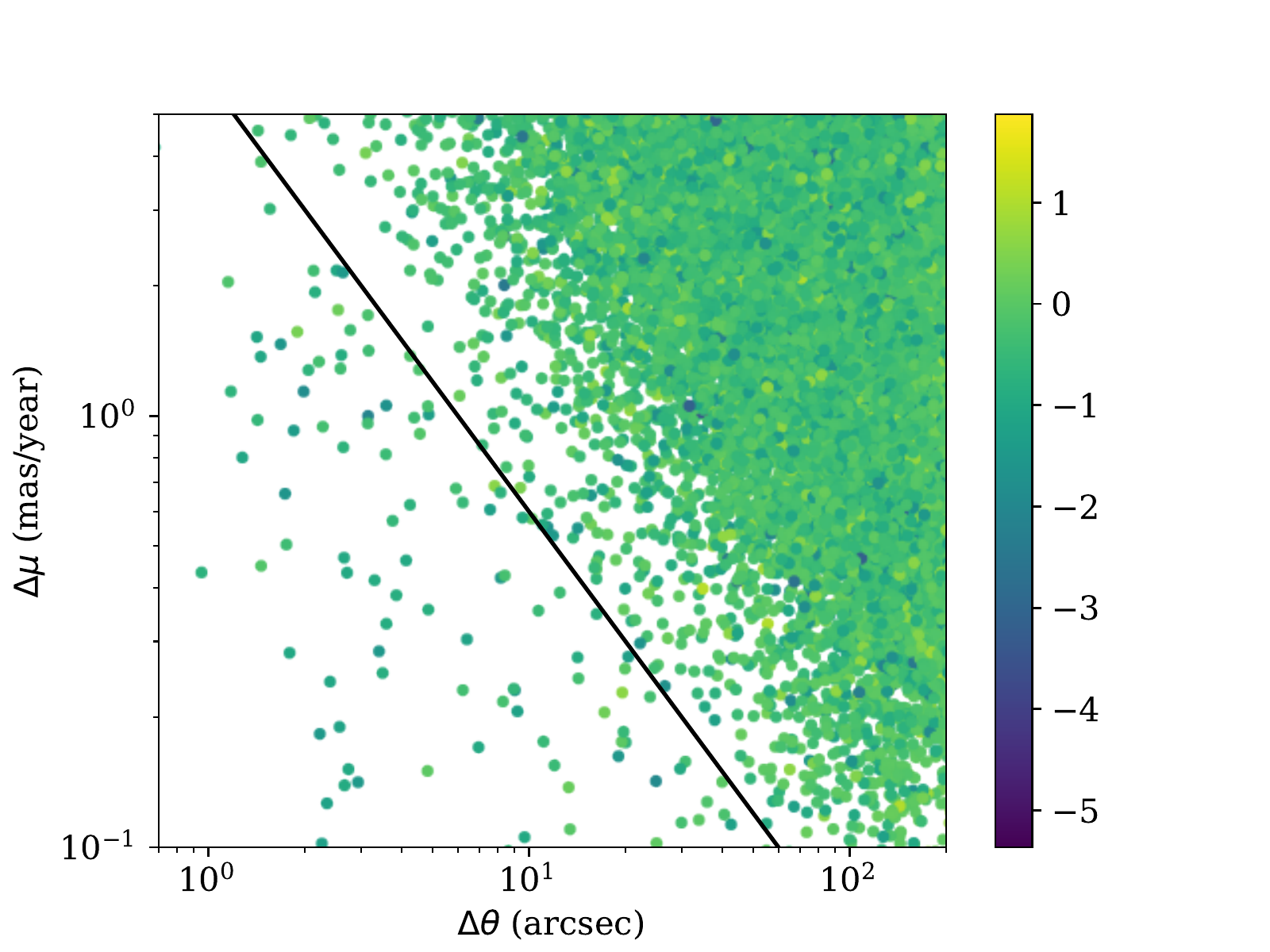}
\end{minipage}
\begin{minipage}{0.48\linewidth}
	\includegraphics[width=\columnwidth]{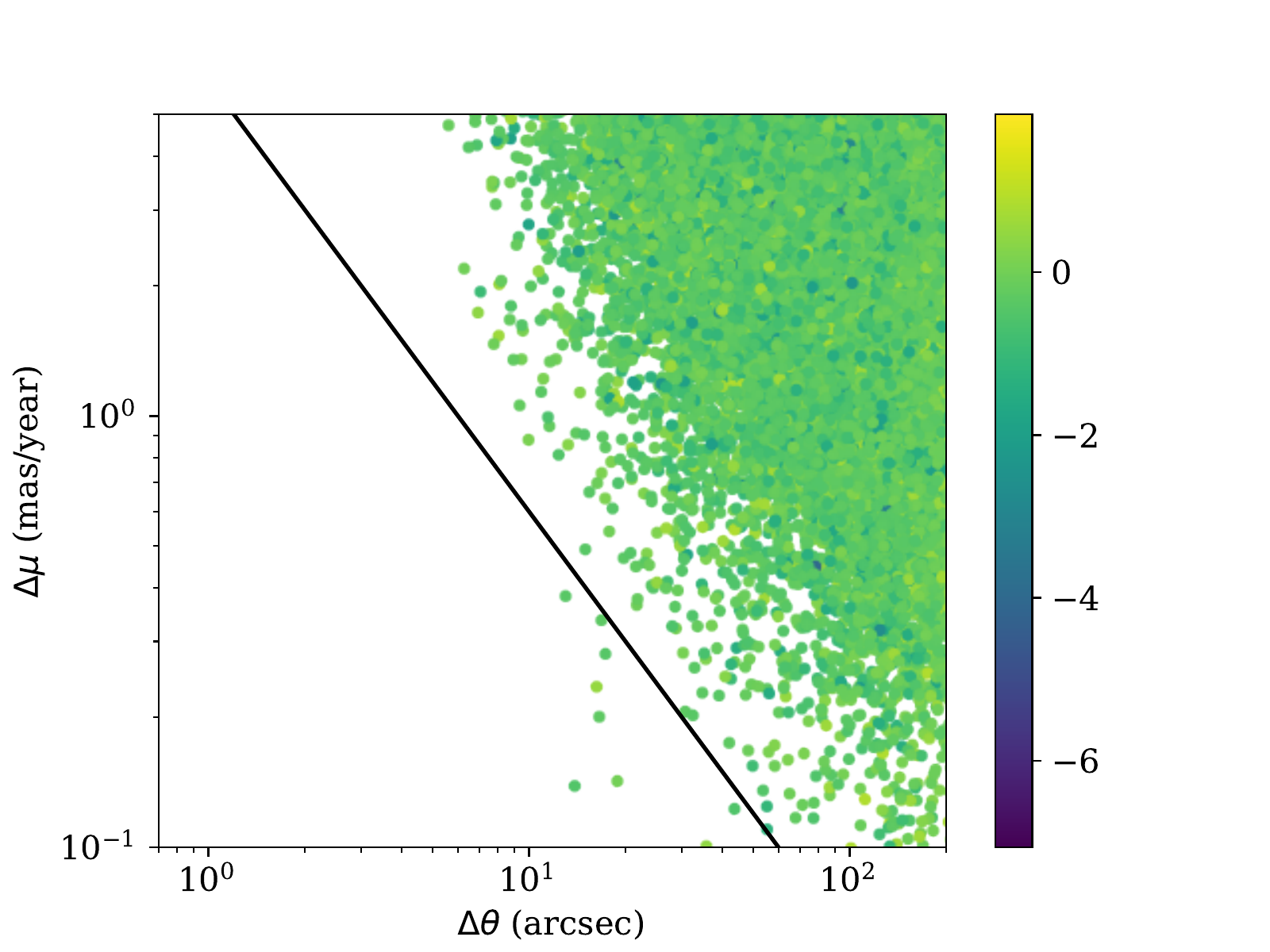}
\end{minipage}
\caption{The difference in proper motions vs. angular separations for sdBs. Left panel: the actual sample. Right panel: all the objects are shifted by 1.4 degree in the direction of right ascension as to exclude any possible real wide binaries from the mock-sample. The color shows the logarithm of the parallax difference. Black solid line assists comparison between the misplaced and observed sample.}
    \label{f:sdb_ang}
\end{figure*}

In our sample we identify 68 ultra-wide binaries for sdB stars, see Table~\ref{t:sdB_candidates}, Figure~\ref{f:sdb_ang} and Table~\ref{t:fract} for multiplicity fractions. It means that sdBs have 3.6 times smaller ultra-wide multiplicity fraction than found in the comparison sample A which is located at similar distances. Another probe is a comparison with MSWD systems which did not go through the CEE. The ultra-wide multiplicity fraction is three times  smaller than in that sample. There is a small caveat in this comparison which we test in length in Appendix~\ref{s:justification}; however we find it can only make up to  1.3 times difference between these fractions.

We plot the cumulative distribution of projected separations for ultra-wide pairs to sdBs in Figure~\ref{f:cum_orb_sep}. On average the ultra-wide companions are located at larger distances than the ones found in the MSWD sample and at smaller distances than one found in  comparison sample A.
The probability for the distributions of projected orbital separations for the wide companions to sdBs and that of MSWDs (i.e. triples) are similar is $2.4\times 10^{-5}$ according to the Kolmogorov-Smirnov (KS) test. The probability that the projected separations for the ultra-wide companions in the comparison sample A (i.e. binaries) and that of the sdBs are drawn from the same distribution is $6\times 10^{-3}$ according to the KS test. 

We also estimated the chance alignment contamination. In order to do so, we shifted the position of each sdB star in the catalogue by 1.4~degree in declination. We then performed the search for ultra-wide binaries using these synthetic positions. We found 8 pairs, two of them paired with the actual host, as it turned out that a few stars are located closer than $\varpi = 10$ and the shift of 1.4~degree is not sufficient to exclude them from the search region. Therefore, the chance alignment contamination is $6/4709\approx 0.0013\pm 0.0005$; this value is ten times smaller than the detected ultra-wide multiplicity for sdBs, affirming that the detected wide companions are likely genuine and are not the result of background contamination.

As an additional check we searched  Table~\ref{t:sdB_candidates} and checked the literature for known close companions for any of these objects. Our original list contained 6 more objects which were excluded as we briefly discuss below. 

$\zeta^1$ Cnc A \citep{1981ApJS...45..437A,1950ApJ...112..554R} is a known resolved binary with a sdB component which we also found in the Gaia. 

BD-12134A is known to be hierarchical triple at the center of the planetary nebula NGC 246 \citep{2014MNRAS.444.3459A}.  BD-12134C is located at separation of $\approx 1$~arcsec from the BD-12134A. In our analysis we could not identify it since star is very red and faint with $J\approx 18.4$.

The system CD-229142 was suspected to be a binary \citep{2000PASP..112..354S}, the Gaia search turned out an additional component to this system.

CD-4214462 seems to be a binary with spectroscopically identified white dwarf \citep{1999ApJS..121....1M}.

CPD-73420 is known to be a binary star \citep{2012yCat.1322....0Z}.

EC21494-7018 might have an extremely low-mass white dwarf companion according to \cite{10.1093/mnras/stv821}.

HD136176B and HD166370B are known to be visually resolved binaries \citep{2001A&A...378..954G, 2012AstL...38..694G}.

PG0834+501 shows variations of radial velocity with amplitude $\approx 50$~km/s \citep{1998ApJ...502..394S, 2005MNRAS.364.1082G}.

TYC6347-931-1 is known to have a visually resolved companion according to the Simbad database. 

V*AHMen is an accreting WD emitting X-ray \citep{1984ApJS...56..507W,2017PASP..129f2001M}.

V*TXCol is an intermediate polar \citep{2019MNRAS.482.3622S,1986ApJ...311..275T} with an orbital period of $\approx 5.7$~hours.

Additionally we noticed that some of sdBs stars actually belong to the globular cluster NGC 6752. In this case multiple stars could be seen as ultra-wide pairs, therefore we removed these objects from our list. Also  V*AHMen and V*TXCol are already present in our list of CVs and therefore we  excluded them from the sdB list.

\begin{figure}
	\includegraphics[width=\columnwidth]{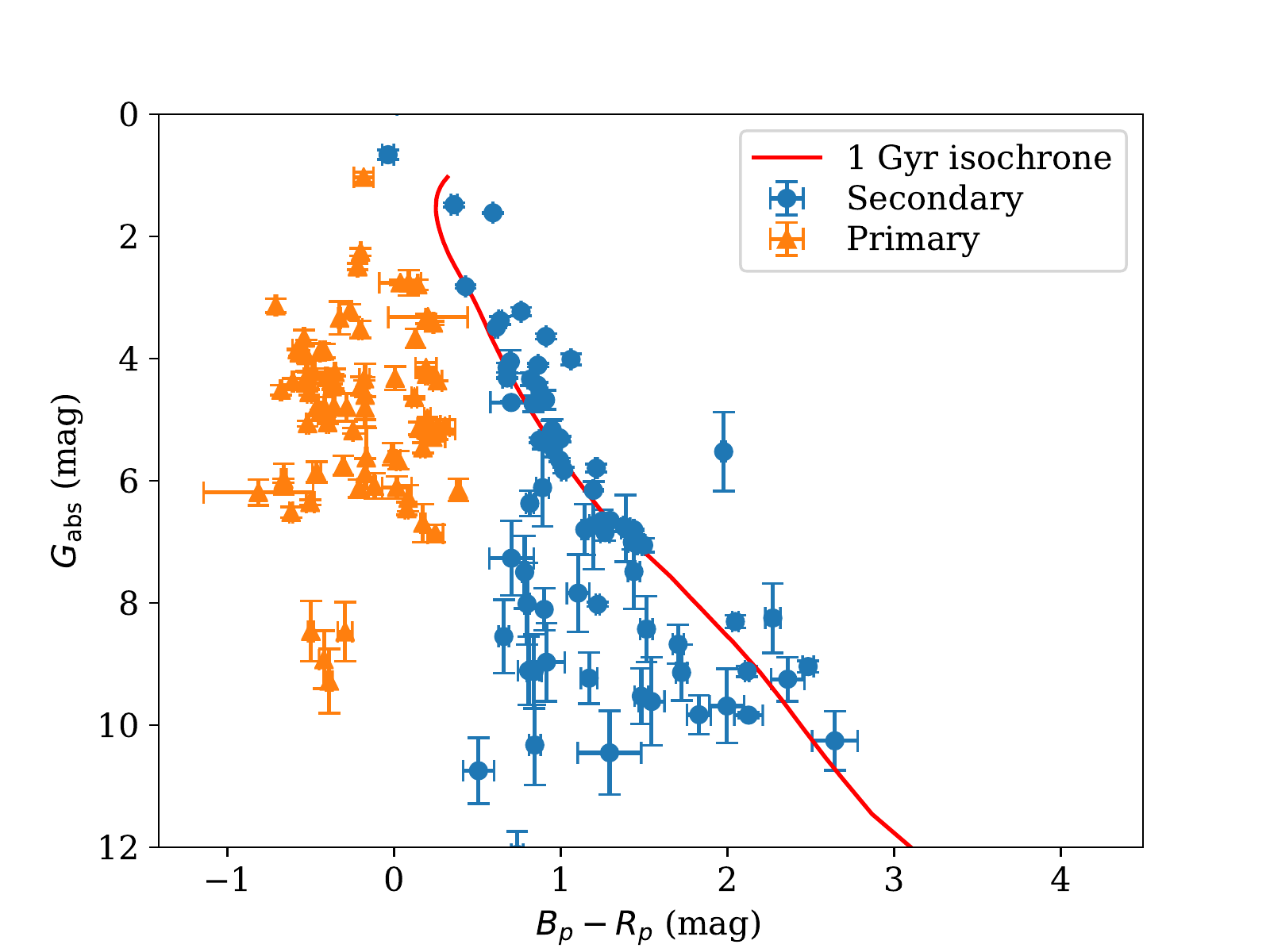}
\caption{The Hertzsprung Russell diagram for ultra-wide companions to sdBs. 
}
    \label{f:hr_mass_sdb}
\end{figure}

Following the prescription from the previous section we have similarly prepared the Hertzsprung Russell diagram for the primary sdB stars and their ultra-wide companions, see Figure~\ref{f:hr_mass_sdb}. The companions seems to be normal typical low-mass main sequence stars. We notice that four primary systems lay close to the top of the white dwarf sequence ($G_\mathrm{abs}\approx 9$ and $B_p - Rp \approx -0.5$) and, therefore, they might in fact be white dwarf and not sdB stars. We therefore excluded the following potential contaminants to the sdB sample: Gaia DR2 1605126585296788480, 5957303154940605696, 2MASSJ14360144+5227424 and Gaia DR2 2867830997336128256.

It is worth noticing that the distribution of projected separations for the ultra-wide companions of MSWD binaries sample is skewed toward smaller separations in comparison with the sdB sample (see Figure~\ref{f:cum_orb_sep}). We believe that this effect occurs because the MSWD are composed of low-mass stars with long timescale of mass ejection (AGB stage) while sdBs are more massive stars with possible fast mass ejection. The fast mass ejection unbinds many utra-wide systems. Systems which survive fast mass ejections tend to be wider than ones surviving slower mass ejection.

\begin{figure*}
    \begin{minipage}{0.49\linewidth}
	\includegraphics[width=\columnwidth]{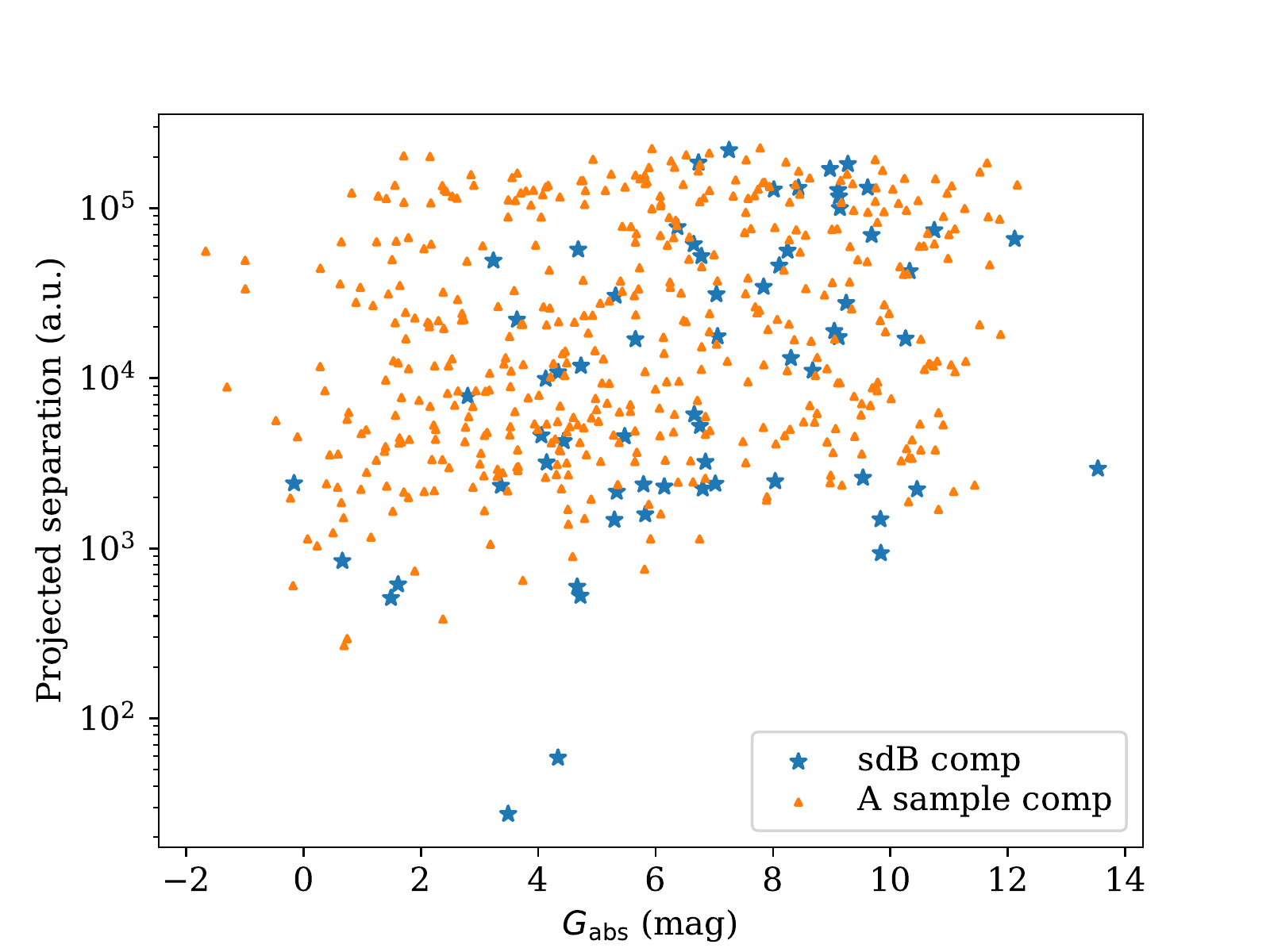}
	\end{minipage}
	\begin{minipage}{0.49\linewidth}
	\includegraphics[width=\columnwidth]{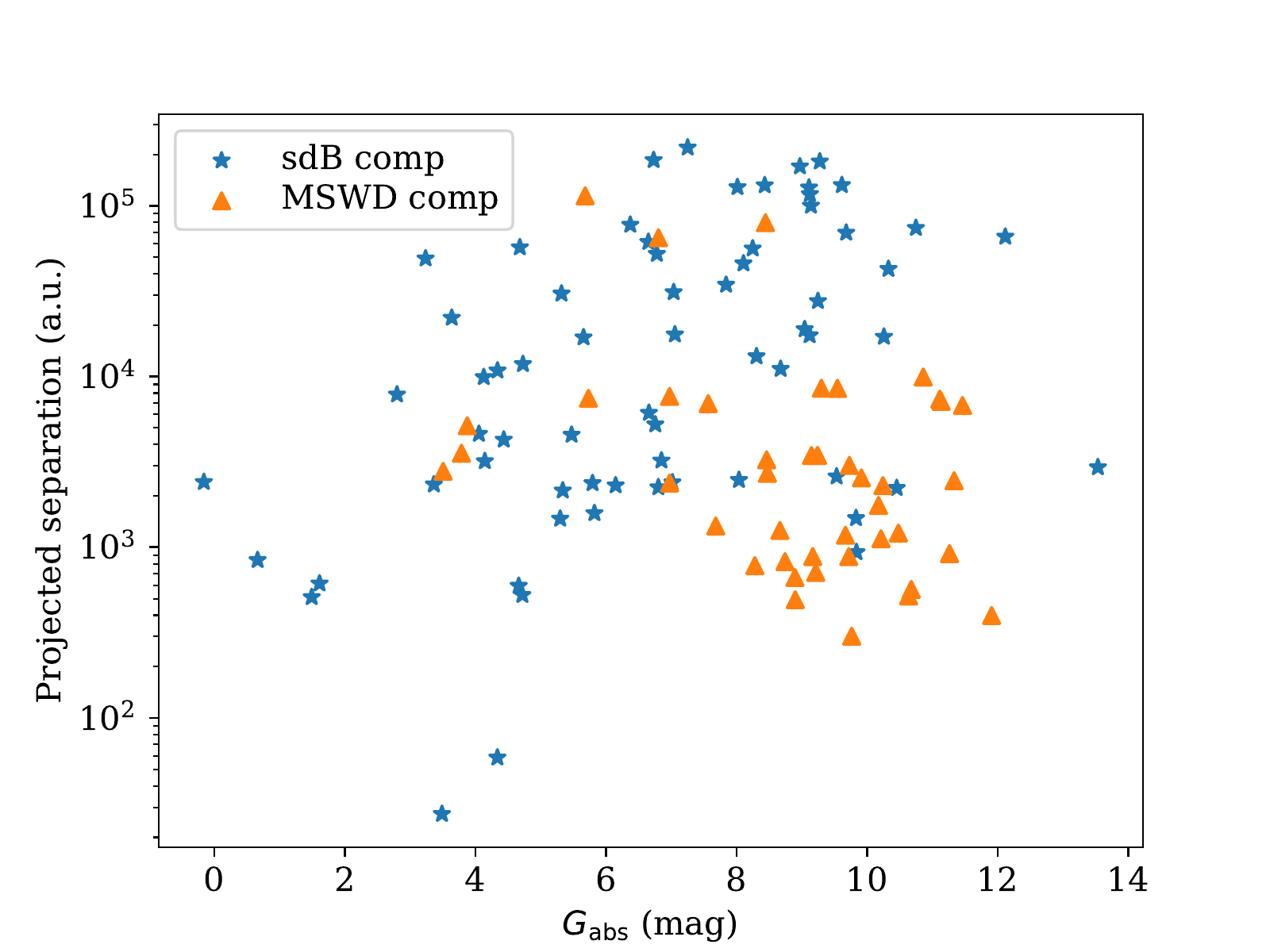}
	\end{minipage}
\caption{Projected separation of the ultra-wide companions as a function of their absolute magnitude for stars in the A sample (left panel) and the MSWD (right panel).}
    \label{f:proj_sep_abs_mag}
\end{figure*}

We also test whether the difference in projected separations between the sdB sample and the comparison sample A is caused by some observation selection effect or whether it is a real effect. To do so we plot the companion absolute magnitude versus projected separation for both samples in Figure~\ref{f:proj_sep_abs_mag} (left panel). We see that sdB's companions cover the same range of projected separations as companions to stars from the A sample with an exception of two known close-by companions to the stars zet01 Cnc and HD136176B.  

If we compare the MSWD sample with sdBs in a similar Figure~\ref{f:proj_sep_abs_mag} (right panel), we notice that the companions to MSWDs are more concentrated towards the fainter side, an aspect which might be partly explained by the smaller masses of the MSWDs in comparison to the sdBs progenitors.

\begin{table*}
    \centering
    \begin{tabular}{llccrcccrccccc}
        \hline
        Name        &  Gaia primary & Gaia secondary & $\Delta \theta$  & $\Delta\varpi \pm \sigma_{\Delta\varpi}$ & $\Delta \mu\pm \sigma_{\Delta\mu}$ & $A$  & Mult. \\
                    & Gaia DR2 name & Gaia DR2 name  & (arcsec)           & (mas)                                          & (mas year$^{-1}$)                        & (a.u.)\\
        \hline
*zet01CncA  &  657244521593509376  &  657244586015485440  &  1.1  & $ 0.339  \pm  0.237 $ & $ 64.592  \pm  0.374  $ &  27.6 & Known \\
2MASSJ01531776+3542049  &  330261789400007040  &  330264916136195072  &  92.0  & $ 0.226  \pm  0.379 $ & $ 0.129  \pm  0.76  $ &  83948.3 & No \\
2MASSJ08251803+1131062  &  601188910547673728  &  601188910547673600  &  2.7  & $ 0.114  \pm  0.071 $ & $ 0.139  \pm  0.116  $ &  3856.5 & No \\
BD-12134A  &  2376592910265354368  &  2376592910265354496  &  3.9  & $ 0.153  \pm  0.106 $ & $ 0.383  \pm  0.22  $ &  1982.9  & Known\\
BPSBS17579-0012  &  2550973079312403712  &  2550973083608089984  &  2.7  & $ 0.104  \pm  0.07 $ & $ 0.432  \pm  0.113  $ &  1686.6 & No \\
CD-229142  &  3492203484216217856  &  3492203484216217728  &  7.6  & $ 0.057  \pm  0.088 $ & $ 0.605  \pm  0.09  $ &  2209.0  & Known\\
CD-4214462  &  6688624794231054976  &  6688624794233492864  &  6.9  & $ 0.16  \pm  0.17 $ & $ 1.206  \pm  0.234  $ &  916.1  & Yes\\
CPD-73420  &  5262163664424670848  &  5262163664426978816  &  1.7  & $ 0.026  \pm  0.046 $ & $ 0.659  \pm  0.093  $ &  529.9  & Known\\
EC05015-2831  &  4877263019073081600  &  4877263023370516096  &  20.9  & $ 0.13  \pm  0.081 $ & $ 0.334  \pm  0.134  $ &  12146.6 & No \\
EC21494-7018  &  6395639996658760832  &  6395639923642870272  &  15.4  & $ 0.316  \pm  0.52 $ & $ 1.231  \pm  0.725  $ &  3132.3  & Yes?\\
FBS0638+428  &  963881581386403072  &  963881577091962368  &  3.8  & $ 0.284  \pm  0.308 $ & $ 0.57  \pm  0.444  $ &  3233.5 & No  \\
Feige91  &  1660055029417965952  &  1660055098137442944  &  39.8  & $ 0.074  \pm  0.102 $ & $ 0.441  \pm  0.165  $ &  18292.1  & No\\
GALEXJ063952.0+515658  &  992534888766785024  &  992534888766784640  &  7.0  & $ 0.1  \pm  0.075 $ & $ 0.17  \pm  0.135  $ &  2384.9 & No \\
GALEXJ110055.9+105542  &  3868418219635118080  &  3868418219635275520  &  8.6  & $ 0.332  \pm  0.172 $ & $ 0.098  \pm  0.458  $ &  8019.4 & No \\
HD136176B  &  1271209615518148736  &  1271209611223823232  &  1.6  & $ 0.059  \pm  0.066 $ & $ 11.352  \pm  0.081  $ &  58.8 & Known \\
HD137737  &  5820064282494545280  &  5820064282512073728  &  2.7  & $ 0.101  \pm  0.069 $ & $ 0.468  \pm  0.104  $ &  2208.9  & No \\
HD166370B  &  6726045641698117888  &  6726045641691326976  &  2.0  & $ 0.015  \pm  0.124 $ & $ 1.136  \pm  0.306  $ &  847.1 & Known  \\
KPD2254+5444  &  2002880555945732992  &  2002880555945731968  &  3.4  & $ 0.032  \pm  0.085 $ & $ 0.284  \pm  0.131  $ &  2347.0  & No\\
M27  &  1827256624493300096  &  1827256628817680896  &  6.4  & $ 0.067  \pm  0.062 $ & $ 0.303  \pm  0.086  $ &  2417.0  & No\\
NGC675250  &  6638380690556259072  &  6638376567387480704  &  244.0  & $ 0.096  \pm  0.435 $ & $ 0.077  \pm  0.495  $ &  182602.9 & No \\
 &  364729314267240192  &  364729314267239936  &  9.2  & $ 0.066  \pm  0.13 $ & $ 0.206  \pm  0.208  $ &  10911.3  \\
  &  340170996210473856  &  340171000507010176  &  2.2  & $ 0.034  \pm  0.196 $ & $ 0.183  \pm  0.225  $ &  1575.4  \\
  &  170774775937432832  &  170774879016640000  &  61.3  & $ 0.394  \pm  0.294 $ & $ 0.079  \pm  0.498  $ &  88246.9  \\
  &  3009110712427319296  &  3009110716723715840  &  42.1  & $ 0.059  \pm  0.07 $ & $ 0.094  \pm  0.13  $ &  46036.1  \\
  &  3342874205845523072  &  3342874240205256704  &  38.1  & $ 0.062  \pm  0.059 $ & $ 0.197  \pm  0.088  $ &  16508.8  \\
  &  196726961201325568  &  196726961201325184  &  4.2  & $ 0.104  \pm  0.105 $ & $ 0.461  \pm  0.16  $ &  2982.7  \\
  &  5280973147283545344  &  5280973525240620928  &  136.3  & $ 0.005  \pm  0.04 $ & $ 0.237  \pm  0.072  $ &  61316.1  \\
  &  5614913348547819392  &  5614913211099174656  &  120.2  & $ 0.336  \pm  0.232 $ & $ 0.061  \pm  0.398  $ &  96390.8  \\
  &  690626278727938304  &  690626278727938560  &  2.4  & $ 0.065  \pm  0.066 $ & $ 0.126  \pm  0.053  $ &  2495.0  \\
  &  5355946268217174656  &  5355946268217174400  &  4.3  & $ 0.015  \pm  0.033 $ & $ 0.095  \pm  0.059  $ &  4334.6  \\
  &  5198534239334516992  &  5198534204974777984  &  19.1  & $ 0.03  \pm  0.064 $ & $ 0.162  \pm  0.112  $ &  17173.0  \\
  &  5856360741911875840  &  5856360776271617024  &  26.7  & $ 0.218  \pm  0.142 $ & $ 0.049  \pm  0.23  $ &  36687.5  \\
  &  6083515098240961920  &  6083515858449657344  &  191.9  & $ 0.2  \pm  0.479 $ & $ 0.069  \pm  0.635  $ &  133618.3  \\
  &  5899284885568130560  &  5899284881258356864  &  30.4  & $ 0.775  \pm  0.402 $ & $ 0.312  \pm  0.782  $ &  30377.7  \\
  &  5903913348492755072  &  5903916333476091264  &  110.7  & $ 0.434  \pm  0.233 $ & $ 0.083  \pm  0.44  $ &  165034.5  \\
  &  5983018743361688064  &  5983018743361687424  &  2.3  & $ 0.055  \pm  0.078 $ & $ 0.102  \pm  0.127  $ &  2216.6  \\
  &  5981257806743601792  &  5981257909822816256  &  42.3  & $ 0.033  \pm  0.11 $ & $ 0.55  \pm  0.243  $ &  17692.2  \\
  &  5832927778318780416  &  5832927640879800064  &  97.5  & $ 0.302  \pm  0.183 $ & $ 0.031  \pm  0.251  $ &  117800.1  \\
  &  5938577715980303616  &  5938578510554279552  &  167.8  & $ 0.371  \pm  0.426 $ & $ 0.115  \pm  0.768  $ &  186934.0  \\
  &  4112632469609705472  &  4112632469562573568  &  1.3  & $ 0.089  \pm  0.116 $ & $ 0.799  \pm  0.27  $ &  974.9  \\
  &  5918752941519223424  &  5918752941519228288  &  26.6  & $ 0.018  \pm  0.081 $ & $ 0.236  \pm  0.14  $ &  22451.8  \\
  &  4501644012800425216  &  4501644012800888320  &  4.5  & $ 0.078  \pm  0.08 $ & $ 0.041  \pm  0.14  $ &  4229.8  \\
  &  6709698863028116096  &  6709698863028118528  &  8.2  & $ 0.036  \pm  0.087 $ & $ 0.42  \pm  0.131  $ &  5127.4  \\
  &  4085168799440497792  &  4085168593282058368  &  47.5  & $ 0.082  \pm  0.082 $ & $ 0.054  \pm  0.138  $ &  52037.3  \\
  &  6632375639082824064  &  6632375673442542720  &  59.9  & $ 0.345  \pm  0.223 $ & $ 0.128  \pm  0.246  $ &  62553.0  \\
  &  6632375639082824064  &  6632372237462530048  &  136.0  & $ 0.874  \pm  0.464 $ & $ 0.051  \pm  0.524  $ &  141949.5  \\
  &  6632375639082824064  &  6632375501638566912  &  179.7  & $ 0.474  \pm  0.393 $ & $ 0.082  \pm  0.444  $ &  187551.0  \\
  &  4301614775823709568  &  4301614737148533504  &  29.7  & $ 0.108  \pm  0.258 $ & $ 0.152  \pm  0.433  $ &  39432.0  \\
  &  1821561467841789312  &  1821561467841789056  &  2.7  & $ 0.103  \pm  0.057 $ & $ 0.151  \pm  0.07  $ &  3629.3  \\
  &  2035560412373115648  &  2035566558436603392  &  261.3  & $ 0.091  \pm  0.283 $ & $ 0.075  \pm  0.447  $ &  194292.8  \\
  &  4299127989744823424  &  4299128058464308096  &  25.0  & $ 0.012  \pm  0.061 $ & $ 0.142  \pm  0.085  $ &  9876.4  \\
  &  2686841281644006656  &  2686841071191184512  &  81.6  & $ 0.124  \pm  0.142 $ & $ 0.064  \pm  0.173  $ &  69282.1  \\
  &  2686841281644006656  &  2686850386972924544  &  182.8  & $ 0.345  \pm  0.254 $ & $ 0.039  \pm  0.345  $ &  155168.3  \\
  &  2208678999172871424  &  2208678999172872704  &  10.8  & $ 0.025  \pm  0.033 $ & $ 0.068  \pm  0.052  $ &  7995.8  \\
  &  1925448205463385344  &  1925448205467420160  &  1.2  & $ 0.215  \pm  0.184 $ & $ 1.136  \pm  0.247  $ &  666.8  \\
PG0834+501  &  1027028630113289600  &  1027028625817982976  &  44.6  & $ 0.259  \pm  0.278 $ & $ 0.319  \pm  0.308  $ &  23857.3  & Yes\\
PNA6633  &  3827045525522912128  &  3827044765316735104  &  1.8  & $ 0.095  \pm  0.108 $ & $ 0.282  \pm  0.196  $ &  1723.6 & No \\
PNA6634  &  5690534730341025408  &  5690534734636923520  &  9.1  & $ 0.025  \pm  0.108 $ & $ 0.231  \pm  0.203  $ &  10527.4  & No\\
TYC2650-973-2  &  2093326416800046848  &  2093326416802711680  &  1.4  & $ 0.081  \pm  0.072 $ & $ 1.52  \pm  0.144  $ &  525.9 & No \\
TYC3533-2439-1  &  2122270063965861504  &  2122269960886645760  &  46.0  & $ 0.027  \pm  0.059 $ & $ 0.073  \pm  0.104  $ &  31778.1 & No \\
TYC4213-1610-1  &  2161688071217417728  &  2161688071217418240  &  3.6  & $ 0.021  \pm  0.033 $ & $ 1.054  \pm  0.069  $ &  615.7 & No \\
TYC4454-1229-1  &  2251716426898467200  &  2251716426895634560  &  1.9  & $ 0.041  \pm  0.093 $ & $ 0.922  \pm  0.205  $ &  565.3  & No\\
TYC6347-931-1  &  6884185998227170304  &  6884185998226359552  &  1.0  & $ 0.193  \pm  0.116 $ & $ 0.433  \pm  0.15  $ &  589.1 & Known \\
V*MMCrA  &  4035877654477649152  &  4035877933682412928  &  181.2  & $ 0.387  \pm  0.372 $ & $ 0.01  \pm  0.463  $ &  176067.8  & No \\
    \hline
    \end{tabular}
    \caption{The identified sdB systems with wide companions. $A$ is the projected orbital separation. The information about any known multiplicity of the sdB star is provided in Mult. column. with several possible options; Yes - known binarity of the sdB, No - no information, Known - the found companion is known. Question mark means what existence of binary component is not confirmed with certainty. 
    }
    \label{t:sdB_candidates}
\end{table*}

\subsubsection{Distance-parallax conversion}
Before discussing the final results it is also  important to verify whether the projected separations we find are physical and are not affected by some sort of a bias. A potential problem could arise from the direct conversion from distance to parallax.



\begin{figure}
\includegraphics[width=\columnwidth]{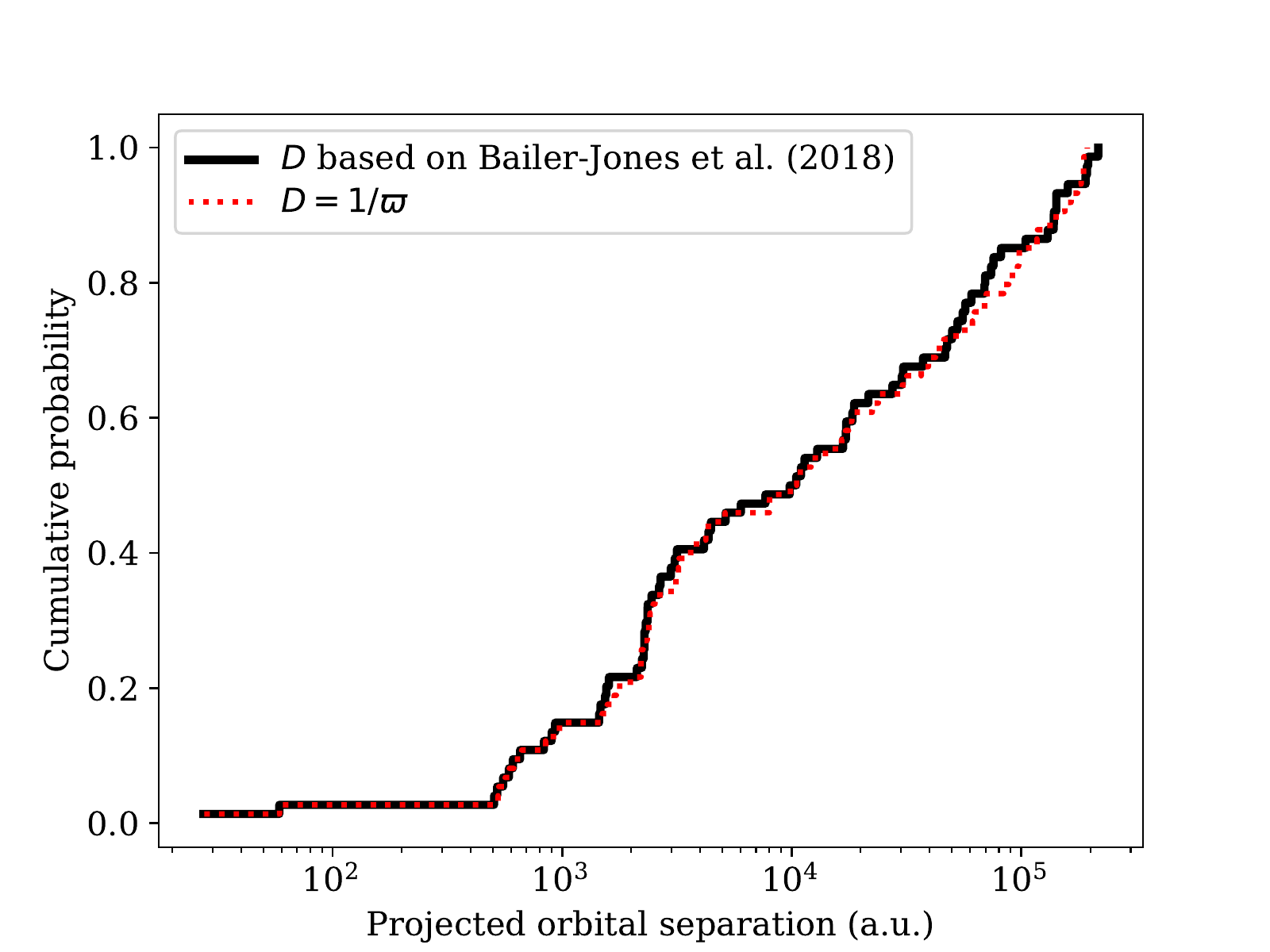}
\caption{The distribution of projected separations for ultra-wide companions to sdB stars derived using $D = 1/\varpi$ (dashed, red line) and the Bayesian estimate for distance based on work by \protect\cite{2018AJ....156...58B} (black, solid line). }
    \label{f:sdb_bailer}
\end{figure}

The conversion from parallax to distance is not straightforward when the accuracy of the parallax measurement is limited \citep{2016A&A...591A.123I,2018AJ....156...58B,2015PASP..127..994B}. To check the contribution of this effect we collect the Bayesian estimates for distances using the catalogue of \cite{2018AJ....156...58B} and plot the cumulative distributions of projected separations in Figure~\ref{f:sdb_bailer}. We find the difference to be negligible.

\subsection{Cataclysmic variables}
We found 14 ultra-wide pairs to cataclysmic variables, see Table~\ref{t:cvs_candidates}. The distribution of the projected separations is shown in Figure~\ref{f:cum_orb_sep}. Half of objects (mostly type NL) are found to have projected separations concentrated around a few$\times 10^3$~a.u., another half (mostly DN) are concentrated at larger separations of a few$\times 10^4$ with the largest separation of J0221+7322 at $\approx 1.6\times 10^5$~a.u.
The multiplicity fraction is two times smaller than the multiplicity fraction found for comparison sample C, see Table~\ref{t:fract}. A smaller multiplicity fraction hints that the systems experienced a more significant mass loss episode. 

Note that one might argue that the comparison of sample C with CVs is invalid, since sample C consists of ultra-wide binaries while CVs with distant companions are in fact triples so their multiplicity fraction might be irrelevant for the comparison. This argument does not seem to be valid because the fraction of CVs with distant companion (2.5~per~cent) is also smaller in comparison to the fraction of MSWD stars with distant ultra-wide companion (i.e. triples) which did not go through a common envelope episode.


\begin{table*}
    \centering
    \begin{tabular}{llccrcccrcccc}
        \hline
        Name        &  Gaia primary & Gaia secondary & $\Delta \theta$  & $\Delta\varpi \pm \sigma_{\Delta\varpi}$ & $\Delta \mu\pm \sigma_{\Delta\mu}$ & $A$  & Type\\
                    & Gaia DR2 name & Gaia DR2 name  & (arcsec)           & (mas)                                          & (mas year$^{-1}$)                        & (a.u.)\\
        \hline
0218+3229  &  325051822271077376  &  325051817976249600  &  5.7  & $ 0.194  \pm  0.343 $ & $ 0.538  \pm  0.647  $ &  2930.3  &  DN  \\
AH Men  &  5207385651533430912  &  5207384891323130368  &  2.9  & $ 0.033  \pm  0.024 $ & $ 0.141  \pm  0.046  $ &  1470.5  &  NL  \\
AY Psc  &  2565601982736199168  &  2565601982736199296  &  39.4  & $ 0.09  \pm  0.12 $ & $ 0.192  \pm  0.198  $ &  29489.7  &  DN  \\
J0154-5947  &  4714563374364671872  &  4714563168206242048  &  8.5  & $ 0.015  \pm  0.046 $ & $ 0.42  \pm  0.076  $ &  2753.3  &  NL  \\
J0221+7322  &  546910213373341184  &  546916569924806272  &  465.5  & $ 0.329  \pm  0.422 $ & $ 0.113  \pm  0.768  $ &  160856.5  &  DN  \\
J0800+1924  &  670132550216853632  &  670132545920724224  &  3.4  & $ 0.666  \pm  0.391 $ & $ 0.331  \pm  0.475  $ &  2352.7  &  DN  \\
J1930+0530  &  4294249387962232576  &  4294249387935557888  &  2.2  & $ 0.102  \pm  0.092 $ & $ 1.395  \pm  0.252  $ &  708.9  &  CV  \\
J2256+5954  &  2014349389931360768  &  2014349389931359616  &  5.9  & $ 0.034  \pm  0.029 $ & $ 0.408  \pm  0.041  $ &  2922.3  &  NL  \\
MR UMa  &  772038105376131456  &  772038105376626432  &  5.5  & $ 0.15  \pm  0.175 $ & $ 0.423  \pm  0.231  $ &  1860.5  &  DN  \\
NGC 104-W1  &  4689639301203677952  &  4689639232475726976  &  39.2  & $ 0.163  \pm  0.383 $ & $ 0.229  \pm  0.519  $ &  20498.7  &  CV  \\
NY Lup  &  5988071549046301184  &  5988071579074013824  &  41.8  & $ 0.121  \pm  0.165 $ & $ 0.055  \pm  0.269  $ &  53197.5  &  NL  \\
TX Col  &  4804695427734393472  &  4804695423438691200  &  2.6  & $ 0.07  \pm  0.044 $ & $ 0.19  \pm  0.087  $ &  2374.8  &  NL  \\
V3885 Sgr  &  6688624794231054976  &  6688624794233492864  &  6.9  & $ 0.16  \pm  0.17 $ & $ 1.206  \pm  0.234  $ &  916.1  &  NL  \\
V453 Nor  &  5984221987022142464  &  5984221987004209920  &  3.2  & $ 0.189  \pm  0.507 $ & $ 1.492  \pm  0.968  $ &  931.1  &  DN  \\
    \hline
    \end{tabular}
    \caption{Wide binaries found in the CVs sample. $A$ is the projected orbital separation. Types are DN - dwarf novae, NL - nova like variable, CV - general CV.}
    \label{t:cvs_candidates}
\end{table*}



\section{Simulations of orbital evolution for systems with common envelope ejection}
\label{s:sim}
In the previous section we demonstrated that the multiplicity fraction for systems which likely went through a CEE is typically 2-4 times smaller than the fractions in the corresponding comparison samples A,B,C or among systems which did not go through the CEE. In this Section we perform gravitational dynamics simulations of  hierarchical triples where we include mass loss from the inner binary in order to estimate the probability for a system to survive the CE ejection episode. We then  compare the resulting projected separations with the ones found in the previous Section.

\subsection{Method}
In our simulations we make a simplified calculation where we replace the primary star in the appropriate comparison sample with a progenitor binary at the CE stage that loses mass at some given rate, where different mass-loss rates are considered as to identify the mass-loss rate that best reproduces the observations. We then follow the evolution of that star and its wide companion, as to synthesize a post-CE-like system with a wide companion that lose mass through the CEE. We follow the evolution of the orbital elements of a distant companion depending on mass loss rate. 

In order to perform the simulations we use a technique similar to the one described in \cite{2019MNRAS.484.4711M} with a small difference. Namely, we consider a ultra-wide binary to be unbound if its orbital energy is positive or its projected separation exceeds the size of our searching region i.e. $2\times 10^5$~a.u.

The initial parameters for our simulations are as follows. First, we select ultra-wide binaries from the comparison sample. Then, we consider five random  eccentricities for each binary sampled from a thermal eccentricity distribution \citep{1937AZh....14..207A,1975MNRAS.173..729H} and five eccentric anomalies from a uniform distribution. While the eccentricity distribution of solar-type binaries is uniform \citep{2010ApJS..190....1R}, the eccentricity distribution of very wide companions is somewhat similar to thermal \citep{2016MNRAS.456.2070T,2017ApJS..230...15M}. We take the semi-major axis to be 1.02 of the observed projected separation based on the analysis by \cite{2011ApJ...733..122D}. 

We assume that the CE ejection starts immediately at the beginning of the simulation. The orbital motion of the system is integrated using the Hermite fourth order integration scheme with addition of a jerk force due to the mass loss \citep{1995ApJ...443L..93H}. The numerical integration continues until the CE ejection is finished i.e. the primary mass reaches $M_\mathrm{final}$. We convert the final masses, orbital positions and velocities into new semi-major axis $a_f$, eccentricity $e_f$ and eccentric anomaly. The final average separation is then computed as:
\begin{equation}
s_f = a_f \left(1 + \frac{1}{2}e_f^2\right)    
\end{equation}
We assume that a binary stays bound after the CE ejection if $s_f < 2\times 10^5$~a.u. and $e_f < 1$.

As we discuss below, each type of system has a different typical total mass-loss, depending on the progenitors and final remnants, and we therefore discuss the simulations results for each type of system individually.

\subsection{Hot subdwarf systems}
We perform the simulations considering two possible final masses; either assuming $M_\mathrm{final}=0.9\, M_\odot$ or $M_\mathrm{final}=0.4\, M_\odot$. We also consider both a constant mass loss rate and an exponentially decaying mass loss rate in form:
\begin{equation}
M (t) = M_0 \exp\left(-t \tau\right)    
\end{equation}
where $M_0$ is the initial mass of the inner binary and $\tau$ is an inverted timescale.
The initial progenitor masses for sdB are between 2.5 and 4~$M_\odot$ (and taken accordingly from samples A and B). Our motivation to choose such massive progenitors is based on study by \cite{2002MNRAS.336..449H,2003MNRAS.341..669H} who suggested that majority of sdB stars are formed from primaries less massive than $3~M_\odot$. The addition of some secondary mass gives us the mentioned upper limit. The final mass could be as small as a mass of a single sdB i.e. $\approx 0.4\, M_\odot$ or an sdB with some low-mass companion i.e. $0.9\, M_\odot$.
We use two samples to simulate the sdBs: (1) using our sample A and (2) using the closer-by systems in sample B.

\begin{figure*}
\begin{minipage}{0.48\linewidth}
\includegraphics[width=\columnwidth]{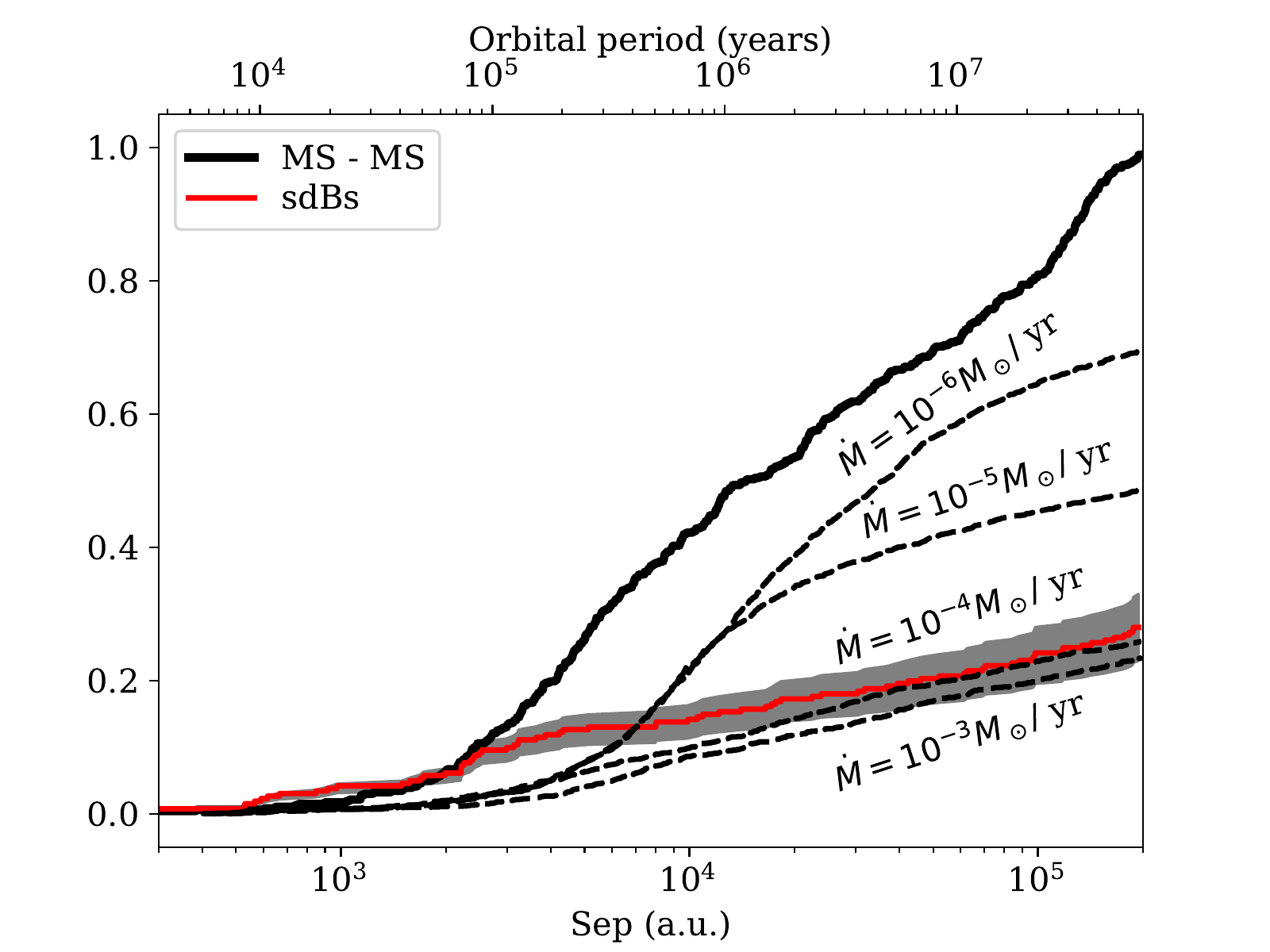}
\end{minipage}
\begin{minipage}{0.48\linewidth}
\includegraphics[width=\columnwidth]{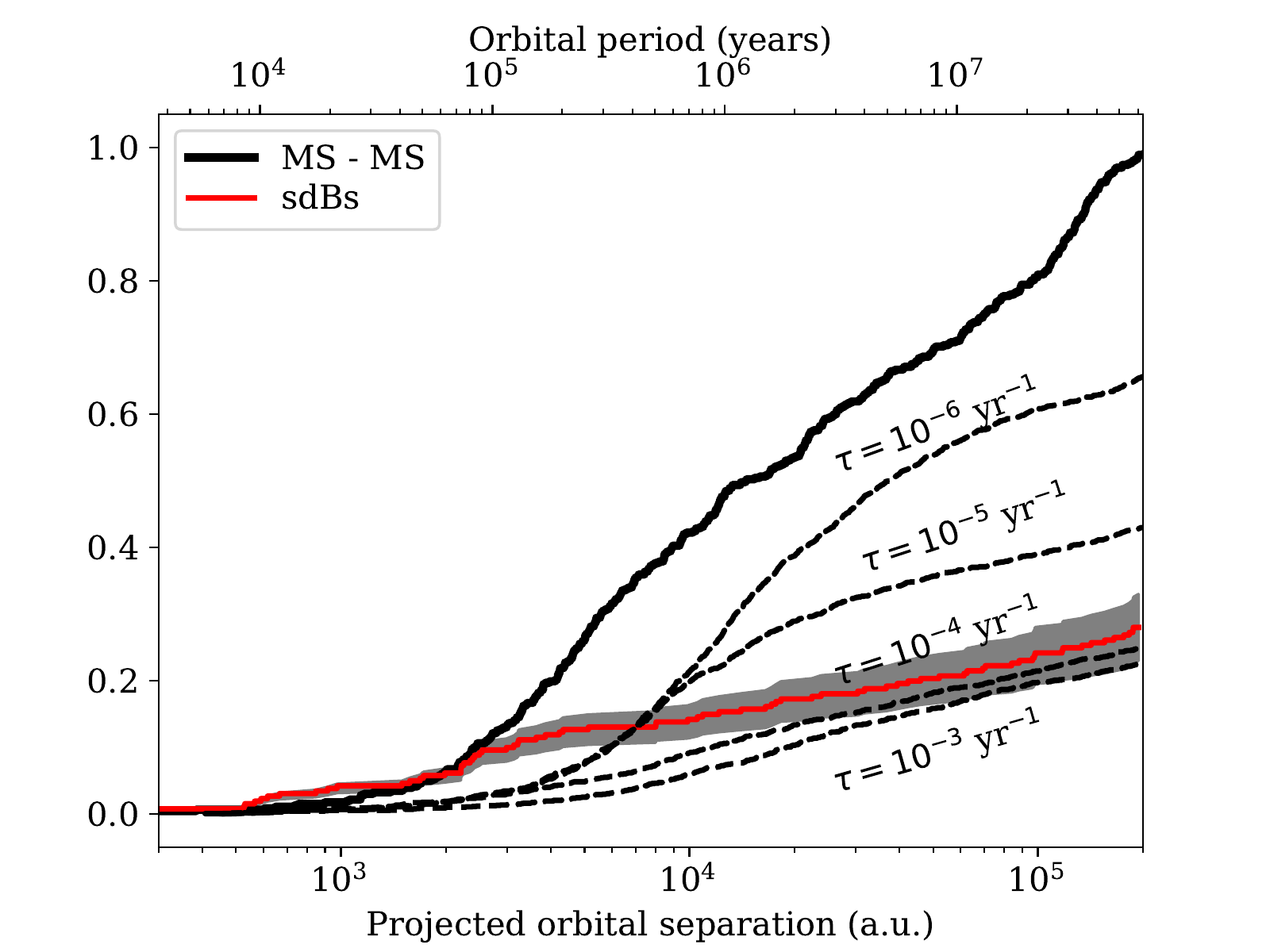}
\end{minipage}
\caption{Results of the simulations of orbital evolution for sdBs with different mass loss timescales using comparison sample A as the initial sample. The final cumulative probability is multiplied by the total survival fraction. Gray area shows the $1\sigma$ uncertainty interval. Left panel - constant mass loss rate, right panel - exponentially decaying mass loss. The final mass of the inner binary is assumed to be $0.9~M_\odot$.}
    \label{f:cum_orb_sep_sim}
\end{figure*}

Using sample A we fail to reproduce the sdBs with ultra-wide companion at separations of $2-4\times 10^3$~a.u., see Figure~\ref{f:cum_orb_sep_sim}.
 We believe this results from the omission of smaller separation systems that can not be resolved in sample A. In particular, systems with $0.5-1\times 10^3$~a.u. separations which are below the Gaia resolution for stars located at distances of $\approx 1$~kpc. Following mass-loss these systems would have widened and fill in the smaller separation regime in the separation distribution. Since these systems are under-sampled in sample A, the resulting simulated systems show a depletion in systems with small separations. In Figure~\ref{f:cum_orb_sep_sim} we scaled the cumulative probability down as to normalize it to the total survival probability computed for the whole sample. Additionally we show the results of our simulations with exponentially decaying mass loss rates in right panel of Figure~\ref{f:cum_orb_sep_sim}. 
 
 In order to overcome the potential problem we performed the same study, but used sample B. This sample of close-by systems better samples even smaller separation systems. 

\begin{figure*}
\begin{minipage}{0.48\linewidth}
\includegraphics[width=\columnwidth]{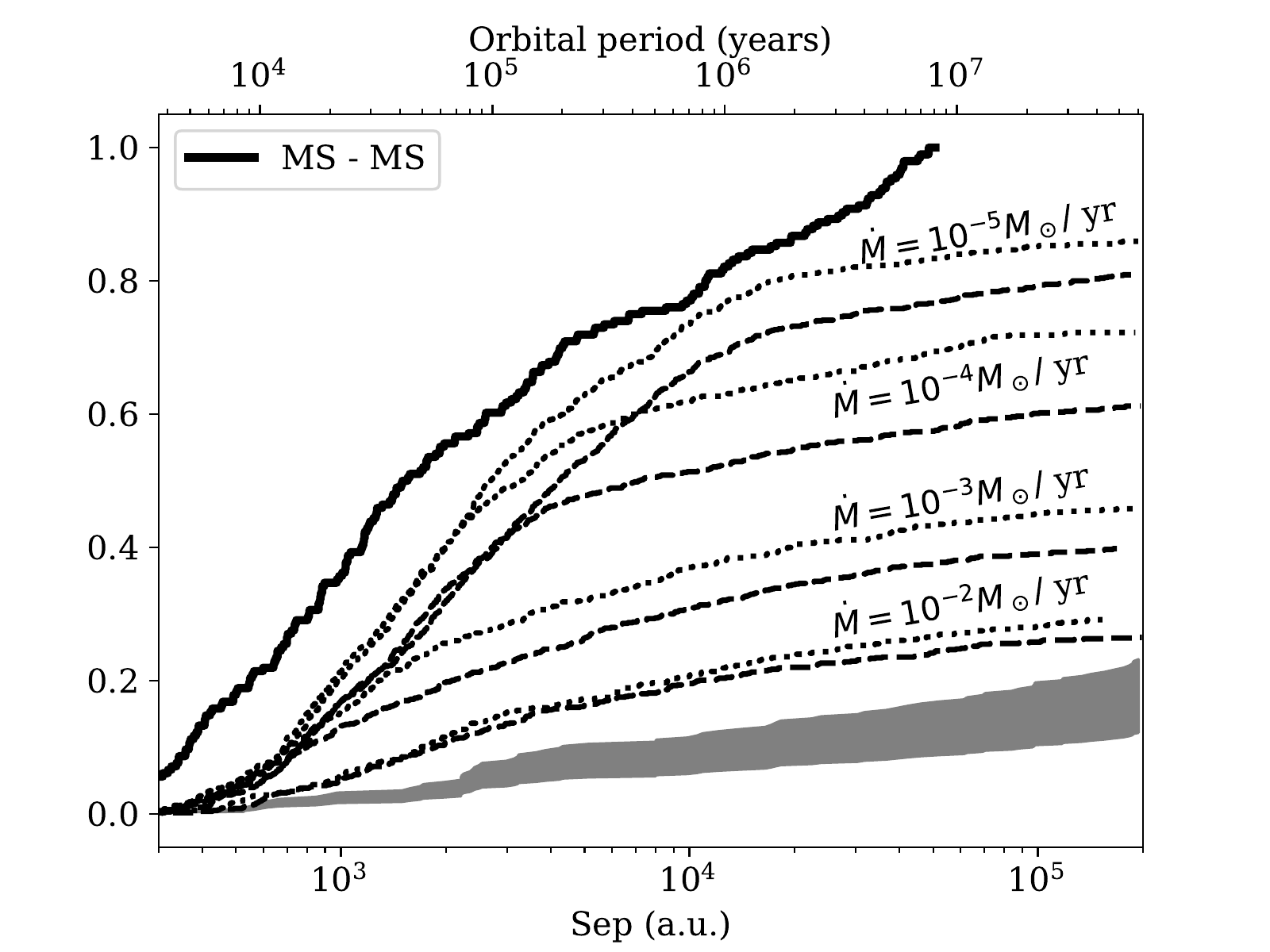}
\end{minipage}
\begin{minipage}{0.48\linewidth}
\includegraphics[width=\columnwidth]{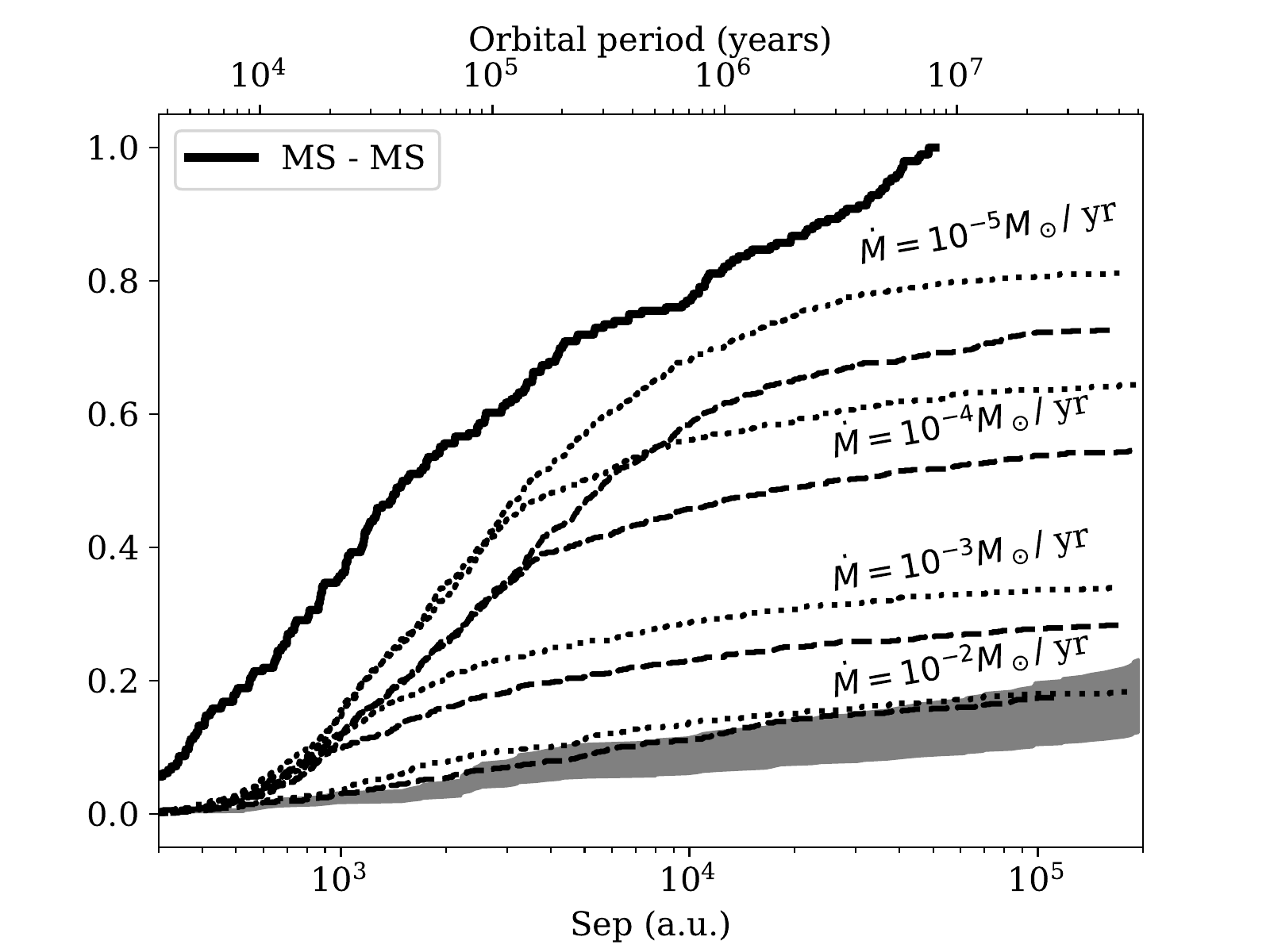}
\end{minipage}
\caption{Results of simulations of the orbital evolution with different mass loss timescales for sdBs using comparison sample B as the initial sample. The left and right panels show simulations with the final mass of the inner binary $0.9~M_\odot$ and $0.4~M_\odot$ respectively.
Dashed and dotted lines are for fractions excluding third components with masses less than $0.4$ and $0.6~M_\odot$ respectively. The grey area shows the uncertainty region for the survival probability of the ultra-wide companions to the sdBs.
The final cumulative probability is multiplied by the total survival fraction. }
    \label{f:sim_better_resol}
\end{figure*}

Using sample B we were able to reproduce the ultra-wide companions at separations of $2-4\times 10^3$~a.u., see Figure~\ref{f:sim_better_resol}. Note, however, that in this case we can not normalize the distribution properly. The close-by stars sample is more sensitive to the detection of fainter companions (below $\sim0.5\, M_\odot$), and therefore can not be directly compared with the large GAIA sample of sdBs.  Nevertheless,  if we set a lower limit of $0.4\, M_\odot$ for the companion we can decrease the ultra-wide binarity fraction and effectively produce a better comparison; in this case the fraction reduced from $\approx 11$~per cent to $\approx 9$~per cent. Even if a larger fraction of sdBs have light ultra-wide binary companions, they are impossible to discover with Gaia at the moment.

Overall, we are able to reproduce the multiplicity of the ultra-wide companions and the distribution of their projected separations only if the CE ejection time scale is compatible with short ejection timescales  i.e. $\dot M \gtrapprox 10^{-2}~M_\odot/\mathrm{year}$. We also tested this by performing additional simplified simulations where the orbital elements were computed using the equation from \cite{1983ApJ...267..322H}.  

\subsection{Cataclysmic variables}
For this simulations we use sample C which includes primary stars with masses in the range $3.5-9\, M_\odot$ and the distant companions with masses larger than $0.4\, M_\odot$. We assume the final mass of the inner binary to be $1.8\, M_\odot$ i.e. there is $\approx 1\, M_\odot$ CO WD and $\approx 0.8\, M_\odot$ secondary star.

\begin{figure}
\includegraphics[width=\columnwidth]{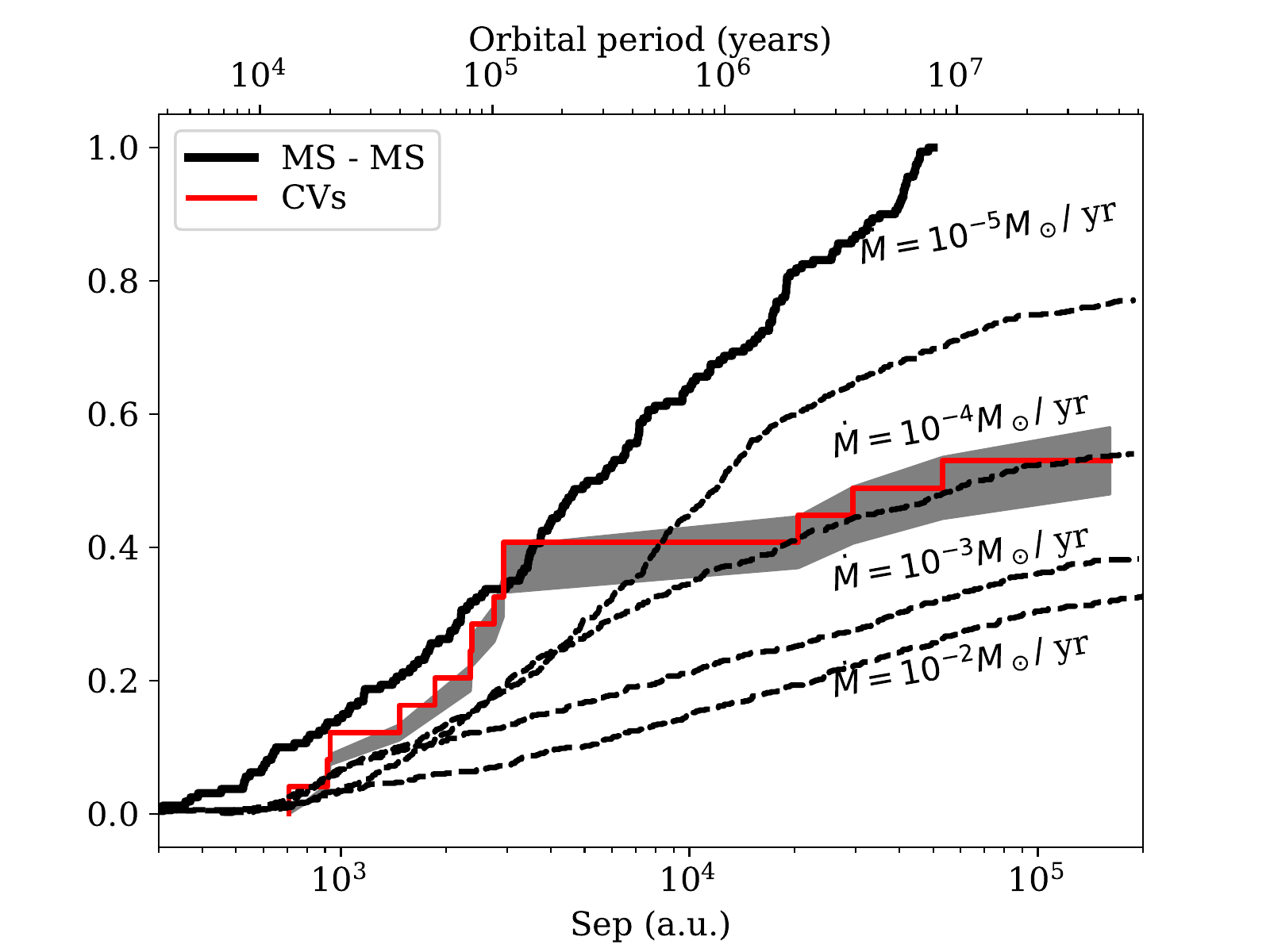}
\caption{Results of the simulations for orbital evolution with constant mass loss rate for ultra-wide binaries with CVs. The final cumulative probability is multiplied by the total survival fraction.}
    \label{f:sim_cv}
\end{figure}

The results of our simulations are shown in Figure~\ref{f:sim_cv}. The small number of CVs with ultra-wide companion is strongly limiting the possibility of a good detailed comparison, and therefore the overall multiplicity fraction is the main indicator for a successful reproduction of the observations. In order to reproduce the observed fraction we find that a longer CE ejection timescale of $\dot M \approx 10^{-4}\, M_\odot/\mathrm{year}$ is required (i.e. a total mass-loss time scale of a few$\times10^4$ yrs is required, given the massive progenitors. 

\section{Discussion and summary}
\label{s:res}
We find that the distributions of projected separations of post-CE systems with additional  wide-companions differ significantly from the distribution of projected separation for ultra-wide companions to corresponding possible progenitor stars in systems which did not go through a CE evolution. We suggest that this can be attributed to the envelope ejection during an episode of CE evolution. In this case the difference in the distributions can be used to constrain the CE process, and in particular the timescale for mass-loss during this process. 

In this work we searched for common proper motion and parallax pairs to systems which went trough a CEE using the Gaia DR 2. We found 68 ultra-wide companions to sdBs, 6 companions to pCE and 14 companions to CVs. Future third Gaia data release will help us to further verify the physical association of these companions through radial velocity measurements. 

We find that the ultra-wide multiplicity rates for systems which went through the common envelope evolution are as follows pCE - 3.7\%, sdBs - 1.4\% and CVs - 2.5 \%. These are 2-4 times smaller than the multiplicity rate found for the corresponding progenitor systems (ultra-wide binaries to wide MSWD - 4.2\%, comparison samples A - 5.0\%, B - 7.0\%, C - 4.7\%). These differences are especially significant in the case of sdBs and CVs.


Assuming that the third companions to sdBs and CVs are bound to the central binary and share a similar physical origin, we perform simulations for the evolution of systems due to mass-loss and consider a range of possible mass loss rates. We find that the fraction of survived ultra-wide companions and the projected separations are compatible to short-term mass loss in the case of sdB formation i.e. $\dot M \gtrapprox 10^{-2}\,M_\odot/\mathrm{year}$. However, in the case of CVs (with the caveat of the much smaller statistics currently exiting), the results suggest much longer timescale of a few $10^4$ yrs (i.e. a mass-loss rate of $\approx 10^{-4}\,M_\odot/\mathrm{year}$). 

Interestingly, studies of the periods of post-CE binary systems gave rise to differences in the inferred $\alpha_{CE}$ parameters between lower and higher mass progenitors \citep{Dav+12}. Though these issues might not be related to our study, they might possibly indicate a joint origin. 
Namely, it is possible that different processes govern CEE in these different systems. For example, it is possible the CEE suggested to be assisted by recombination is sufficiently efficient for low mass-stars below 3 M$_{\odot}$, but less effective for more massive 5-9 M$_{\odot}$ stars (private comm. with P. Podsiadlowski). In this case the more massive progenitors of CVs would not lose their envelope through the inspiral and following phases, and might require a much longer timescale for mass-loss through other means, e.g. through the suggested dust-driven winds mechanism \citep{2018MNRAS.478L..12G} which operates on longer-timescales more consistent with those we inferred for CVs. Such differences would significantly affect the inferred CE parameters.

Finally, CEE might be accompanied by an effective kick to the CE-system due to asymmetric mass-loss. Kicks at the level of even just a few km s$^{-1}$ could dissociate or significantly change the distribution of third wide-companions to such systems. Such possibility would manifest itself as producing smaller fractions of wide companions and at larger separations. It is therefore possible that kicks can mimic the effects of fast mass-loss rate. In the case of sdBs for which the inferred mass-loss rate was high might be alternatively interpreted as a possible evidence for CE-kick, rather that a dynamical mass-loss in CEE. This issue, however is beyond the scope of this study and will be explored in detail elsewhere.  


\section*{Acknowledgements}
We would like to thank P. Podsiadlowski and M. Moe for helpful discussions. A.I.P. thanks the STFC for research grant ST/S000275/1. This work has made use of data from the European Space Agency (ESA) mission
{\it Gaia} (\url{https://www.cosmos.esa.int/gaia}), processed by the {\it Gaia}
Data Processing and Analysis Consortium (DPAC,
\url{https://www.cosmos.esa.int/web/gaia/dpac/consortium}). Funding for the DPAC
has been provided by national institutions, in particular the institutions
participating in the {\it Gaia} Multilateral Agreement.

This research has made use of the SIMBAD database,
operated at CDS, Strasbourg, France


\bibliographystyle{mnras} 
\bibliography{pce_bibl}




\appendix

\section{Identification of MSWD in the Gaia data release}
\label{s:identific}
Many MSWD systems are observed as faint objects with the SDSS g band magnitudes in the range $16^\mathrm{m}-22^\mathrm{m}$, while the Gaia is expected to be complete only until 20 mag in unfiltered light \citep{2004Ap&SS.291..321N,2006MNRAS.367..290J}. A significant fraction of these systems are below the photometric sensitivity of the Gaia. Moreover, in crowded fields the Gaia database could contain up to tens of faint stars in a region with size of a few arcsec. Therefore, we decided to first identify our MSWD binaries in the Gaia database based on both location and magnitude.

To practically search for counterparts, we convert the SDSS colors g,i,r to the G Gaia color using the polynomial fit by \cite{jordi2010}. After this, we select all stars at angular separations less than 5~arcsec from the SDSS catalogue location and assume that the MSWD counterpart is the star which magnitude differs by less than 4$\sigma$ from the magnitude computed according to equation from \cite{jordi2010} and is located at the smallest angular separation from its SDSS catalogue position.

\begin{figure*}
\begin{minipage}{0.48\linewidth}
	\includegraphics[width=\columnwidth]{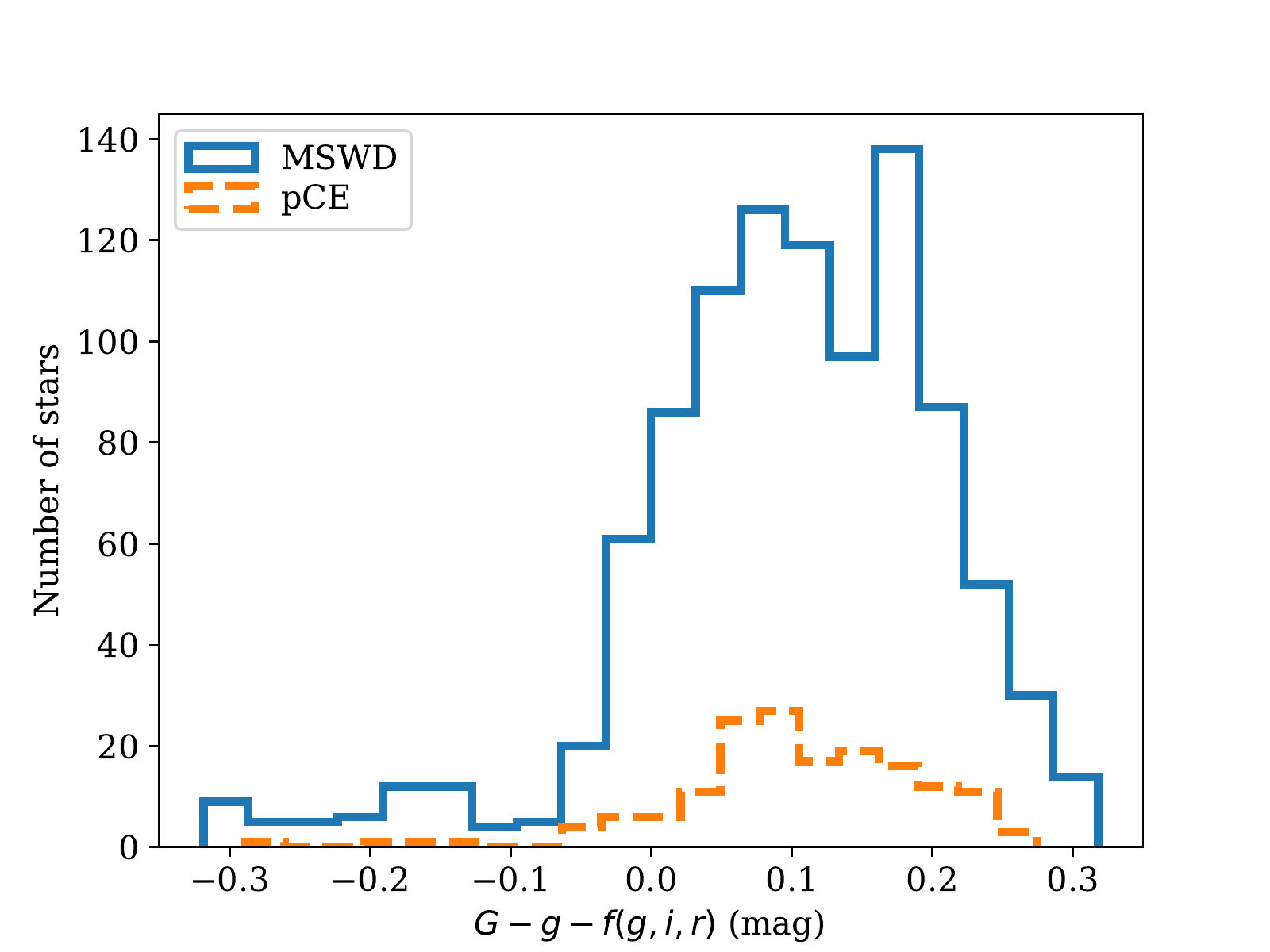}
\end{minipage}
\begin{minipage}{0.48\linewidth}
	\includegraphics[width=\columnwidth]{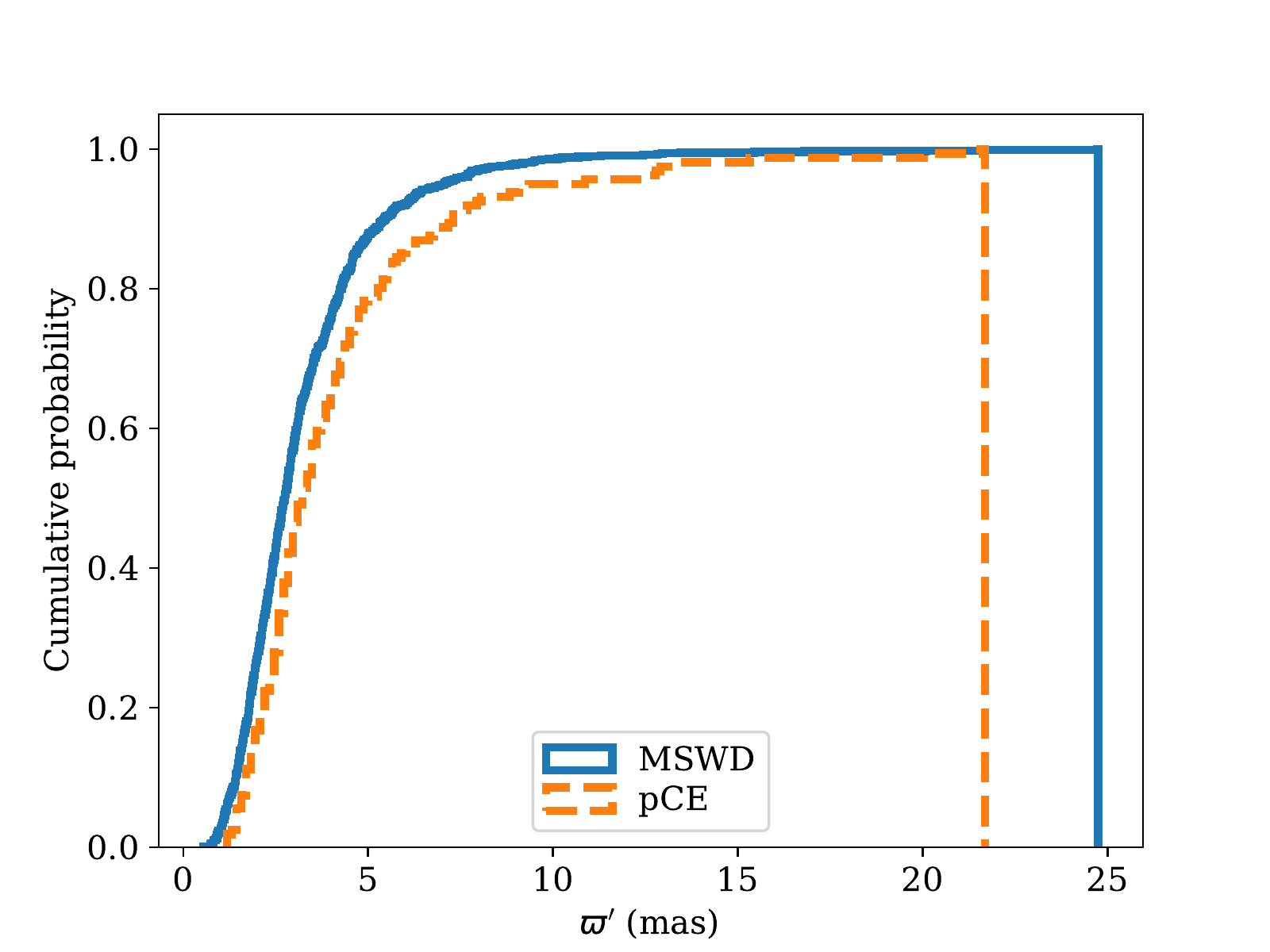}
\end{minipage}
\caption{Left panel: the histogram of the color difference between that predicted by \protect\cite{jordi2010} equation and that measured for Gaia counterpart for general MSWD systems (solid line) and pCE systems (dashed line). Right panel: the cumulative distribution of measured parallaxes for systems identified in Gaia DR2.}
    \label{f:nswd_par}
\end{figure*}

After this we manage to identify 1979  out of 3287 MSWD binaries.
We plot the distribution of color difference and parallaxes in Figure~\ref{f:nswd_par}. We know that in the second Gaia data release stars are treated as separate if the angular separation exceeds a couple of arcsec. For our typical parallax it would correspond to $\approx 600$~a.u. In the real sample the shortest angular separation is $\approx 100$~a.u. which corresponds to an angular separation of 1.2~arcsec. There is no doubt that such systems are seen as spectral binaries in the SDSS survey. In the context of our analysis it means that a fraction of these systems are binaries and not triples. 
 
From the list of systems we choose ones with good astrometric solution and with relative errors of parallax and proper motions smaller than 0.25. Our filtered list contains 998 NSWD binaries including 161 pCE binaries. These binaries have mean parallax 3.3~mas and mean Gaia G magnitude of 18.1. 

\section{ADQL request to select stars for comparison samples}
\label{a:adql_compar}
Here we show two ADQL requests for the Gaia database which helped us to form the comparison samples. The comparison sample A (larger distances) is selected as:
\begin{verbatim}
SELECT top 10000 source_id, ra, dec, phot_g_mean_mag, 
parallax, parallax_error, pmra, pmra_error, pmdec, 
pmdec_error, phot_bp_mean_mag, phot_rp_mean_mag, 
teff_val, lum_val, radius_val 
FROM gaiadr2.gaia_source 
WHERE lum_val > 24 and lum_val < 140 and teff_val > 7500
and teff_val < 10000 and radius_val > 1.4 
and radius_val < 4 and parallax > 0.67 
and parallax < 10 and parallax / parallax_error > 5 
and pmra / pmra_error > 5 and pmdec / pmdec_error > 5 
ORDER by source_id
\end{verbatim}

In order to select the comparison sample B (smaller distances) we use the following request:
\begin{verbatim}
select top 5000 source_id, ra, dec, phot_g_mean_mag,
parallax, parallax_error, pmra, pmra_error, pmdec, 
pmdec_error, phot_bp_mean_mag, phot_rp_mean_mag, 
teff_val, lum_val, radius_val, bp_rp 
from gaiadr2.gaia_source 
where phot_g_mean_mag 
- 5.0 * log10(100.0 / parallax) < 1.2 
and bp_rp < 0.35 
and parallax > 5 and parallax / parallax_error > 20 
order by source_id
\end{verbatim}
This request returns only 2452 stars.

Comparison sample C is selected using the following request:
\begin{verbatim}
select top 5000 source_id, ra, dec, phot_g_mean_mag, 
parallax, parallax_error, pmra, pmra_error, pmdec, 
pmdec_error, phot_bp_mean_mag, phot_rp_mean_mag, 
teff_val, lum_val, radius_val, bp_rp 
from gaiadr2.gaia_source 
where phot_g_mean_mag 
- 5.0 * log10(100.0 / parallax) < 0.0 
and bp_rp < 0.35 
and parallax > 2 and parallax / parallax_error > 10 
order by source_id    
\end{verbatim}

\section{Justification for the comparison between different fractions}
\label{s:justification}

MSWD inner binaries are wider in comparison to sdBs progenitors, which could potentially affect our conclusion. The exact reason is that the observed period distribution peaks around 50 a.u. \citep{tokovininI, tokovininII} and so MSWD binaries in hierarchical triples should be more numerous than the much tigher progenitors of sdB binaries in hierarchical triples. We test this possibility using the simulation procedure described by \cite{tokovininII}. Namely, we draw 480000 stellar systems with primary mass in the range 0.85 and 1.5~$M_\odot$ following the Salpeter initial mass function \citep{salpeter}. We classify a binary at any level of hierarchy as pre-sdB if its orbital separation is between 0.01 and 2 a.u. and as MSWD if its orbital separation is between 10 and 200~a.u. (systems with separations 2-10~a.u. could go through CEE). After this classification step we count separately systems which are members of wide (tertiary at separation larger than $10^3$~a.u.) hierarchies and ones which are simply binaries. This procedure is very similar to that done in our research procedure. 

As a result of this analysis, we identified 4398 pre-sdBs with tertiary companion among 66234 pre-sdBs with or without an additional companion which corresponds to roughly 6.6 per cent of all sdBs to be members of wide hierarchical systems.  The same analysis for MSWD gives us 6972 MSWD are bound with ultra-wide tertiary companion among 88897 MSWD with and without tertially companion which is 7.8 per. Although 6.6 per cent differs from 7.8 per cent, the difference is only 1.2 times. Additional cuts on the ultra-wide system mass ratio $Q = M2/M1 > 0.4$ slightly increases the difference up to 1.3 times. Therefore this effect cannot explain the observed difference of 3 times seen between ultra-wide mupliplicity fraction of MSWD and sdB systems.

We compute similar values for our sample A. Namely, we classify a star as an ultra-wide binary if its orbital separation is above $10^3$~a.u. In our simulations we found 42176 such wide binaries and  437824 are isolated which includes 181596 hierarchies with separations less than $10^3$~a.u. The latter hierarchical systems are treated as isolated stars in our analysis using the Gaia because they can not be resolved. These numbers correspond to 8.8 per cent of the ultra-wide multiplicity fraction.
 This fraction is 1.3 times larger than the corresponding fraction of sdBs with ultra-wide companion and it is 1.1 times larger than the fraction of MSWD with wide companion in comparison to the MSWD without any companion which is in perfect agreement with the small difference seen in Table~\ref{t:fract}.
Overall, the stellar statistics could explain the difference of 1.3 times between the ultra-wide multiplicity fraction for A sample and the sdB sample while the difference seen in observations is 3.6 times.

The same can be shown in a more analytical way. We restrict the discussion up to quadruples. If we use the similar notation as \cite{tokovininII} and designate a binary with the longest orbital period (possibly with more hierarchy levels) as L1, and the inner binary around the primary as L11 and the inner binary around the secondary as L12, we can additionally introduce L11+ as triples with ultra-wide companion at separations larger than $10^3$~a.u. and L11- as triples with ultra-wide companion at separations smaller than $10^3$~a.u., so $N^\mathrm{L11} = N^\mathrm{L11+} + N^\mathrm{L11-}$.
We can describe the number of sdBs with ultra-wide companion at separations larger than $10^3$~a.u. as $N_\mathrm{sdBs}^\mathrm{L11+}$+$N_\mathrm{sdBs}^\mathrm{L12+}$ which is an integral of the probability density for the periods $f(P)dP$ over a range of values allowing a formation of sdBs, taking into account the formation of a triple. The similar number of MSWDs with ultra-wide companion is $N_\mathrm{MSWD}^\mathrm{L11+}$+$N_\mathrm{MSWD}^\mathrm{L12+}$. The fraction presented in Table~\ref{t:fract} refers to the number of $N_\mathrm{sdBs}^\mathrm{L11+}$+$N_\mathrm{sdBs}^\mathrm{L12+}$ among all sdBs which is the sum of $N_\mathrm{sdBs}^\mathrm{L1}$ (truly isolated sdBs) and $N_\mathrm{sdBs}^\mathrm{L11}$+$N_\mathrm{sdBs}^\mathrm{L12}$ (resolved and non-resolved triples). So the fraction ``sdBs + distant" is:
\begin{equation}
f_\mathrm{sdBs} = \frac{N_\mathrm{sdBs}^\mathrm{L11+}+N_\mathrm{sdBs}^\mathrm{L12+}}{N_\mathrm{sdBs}^\mathrm{L1} + N_\mathrm{sdBs}^\mathrm{L11}+N_\mathrm{sdBs}^\mathrm{L12}}    
\end{equation}
A similar fraction of ``MSWD + distant" is:
\begin{equation}
f_\mathrm{MSWD} = \frac{N_\mathrm{MSWD}^\mathrm{L11+}+N_\mathrm{MSWD}^\mathrm{L12+}}{N_\mathrm{MSWD}^\mathrm{L1} + N_\mathrm{MSWD}^\mathrm{L11}+N_\mathrm{MSWD}^\mathrm{L12}}    
\end{equation}
A fraction of these equations gives:
\begin{equation}
\frac{f_\mathrm{sdBs}}{f_\mathrm{MSWD}} = \frac{ N_\mathrm{sdBs}^\mathrm{L11+}+N_\mathrm{sdBs}^\mathrm{L12+} }{ N_\mathrm{MSWD}^\mathrm{L11+}+N_\mathrm{MSWD}^\mathrm{L12+} } \frac{ N_\mathrm{MSWD}^\mathrm{L1} + N_\mathrm{MSWD}^\mathrm{L11}+N_\mathrm{MSWD}^\mathrm{L12} }{ N_\mathrm{sdBs}^\mathrm{L1} + N_\mathrm{sdBs}^\mathrm{L11}+N_\mathrm{sdBs}^\mathrm{L12} }
\end{equation}
The first term gives $\approx 0.5$ because MSWD are wider and cover the peak of the period distribution, but due to exactly the same reason the second term gives $\approx 2$ so the total fraction is close to 1. 

There are two fundamental reasons why we find such a small difference in statistics. First, the simulation procedure suggested by \cite{tokovininII} assumes that the period distributions at different levels of hierarchy are independent of each other as soon as the levels are well separated (dynamical truncation factor is negligible). MSWD and sdBs well satisfy this assumption because their discoverable ultra-wide companions are located at separations larger than $10^3$~a.u. Second, the multiplicity fraction of binaries as compared to isolated stars and triples as compared to binaries are very similar and assumed to be 0.466 by \cite{tokovininII}.

It is interesting to note that if the multiplicity fractions at different levels of hierarchy start to differ e.g. if we interested to reproduce fraction of quadruples precisely, it affects $f_\mathrm{sdBs}/f_\mathrm{A}$ making it slightly larger 1.8 (no observational selection, taking into account sdBs around the secondary star) and $f_\mathrm{sdBs}/f_\mathrm{A} = 1.6$ (cut at $Q > 0.4$). Fraction $f_\mathrm{sdBs}/f_\mathrm{MSWD}$ stays close to 1.2. On the other hand, if we want to take into account more realistic criterion for orbital stability \citep{mardling_aarseth}, we get $f_\mathrm{sdBs}/f_\mathrm{MSWD} \approx 0.9$ and  $f_\mathrm{sdBs}/f_\mathrm{A} = 1.4$. In this case we do not reproduce the stellar statistics, because the method suggested by \cite{tokovininII} mimics this criterion only approximately. Therefore, we expect that newer works on stellar statistics might slightly change the conclusion of our work, but we do not expect $f_\mathrm{sdBs}/f_\mathrm{MSWD}$ or $f_\mathrm{sdBs}/f_\mathrm{A}$  differ from 1 too significantly.

\bsp	
\label{lastpage}
\end{document}